\documentclass[12pt,a4paper]{article}

\usepackage{amsmath,amsthm,amssymb,amscd,a4wide}

\usepackage{dsfont}
\usepackage{mathrsfs}
\usepackage{setspace}

\newcommand{\ud}{\mathrm{d}}
\newcommand{\ui}{\mathrm{i}}
\newcommand{\ue}{\mathrm{e}}
\newcommand{\GL}{\mathrm{GL}}
\newcommand{\M}{\mathrm{M}}

\newcommand{\vl}{\boldsymbol{l}}
\newcommand{\valpha}{\boldsymbol{\alpha}}
\newcommand{\vbeta}{\boldsymbol{\beta}}
\newcommand{\gz}{{\mathbb Z}}
\newcommand{\rz}{{\mathbb R}}
\newcommand{\nz}{{\mathbb N}}
\newcommand{\kz}{{\mathbb C}}
\newcommand{\eins}{{\mathds 1}}
\newcommand{\sm}{\kz\setminus[\pm\ui\sigma(L)\setminus\{0\}]}

\newcommand{\smu}{\kz\setminus[\ui\sigma(L)\cup\{0\}]}
\newcommand{\smz}{\kz\setminus[\pm\ui\sigma(L)]}
\newcommand{\dk}{\frac{\ud}{\ud k}}
\newcommand{\lmp}{\lambda^+_{\min}}
\newcommand{\lmm}{\lambda^-_{\min}}

\newcommand{\cP}{\mathcal{P}}

\newcommand{\cL}{\mathcal{L}}

\DeclareMathOperator{\im}{Im}
\DeclareMathOperator{\re}{Re}
\DeclareMathOperator{\tr}{Tr}
\DeclareMathOperator{\mtr}{tr}
\DeclareMathOperator{\ran}{ran}

\DeclareMathOperator{\arth}{artanh}
\DeclareMathOperator{\erfc}{erfc}

\numberwithin{equation}{section}

\newtheorem{theorem}{Theorem}[section]
\newtheorem{lemma}[theorem]{Lemma}
\newtheorem{prop}[theorem]{Proposition}
\newtheorem{cor}[theorem]{Corollary}
\theoremstyle{definition}
\newtheorem{defn}[theorem]{Definition}
\newtheorem{rem}[theorem]{Remark}

\begin{document}

\thispagestyle{empty}

\noindent

\vspace*{1cm}

\begin{center}

{\LARGE\bf The trace formula for quantum graphs with\\[5mm]
general self adjoint boundary conditions} \\
\vspace*{3cm}
{\large Jens Bolte}%
\footnote{E-mail address: {\tt jens.bolte@rhul.ac.uk}}

\vspace*{6mm}

Department of Mathematics\\
Royal Holloway, University of London\\
Egham, TW20 0EX, United Kingdom\\

\vspace*{1cm}

{\large Sebastian Endres}%
\footnote{E-mail address: {\tt sebastian.endres@uni-ulm.de}}

\vspace*{6mm}

Institut f\"ur Theoretische Physik\\
Universit\"at Ulm, Albert-Einstein-Allee 11\\
D-89069 Ulm, Germany

\end{center}

\vfill

\begin{abstract}
We consider compact metric graphs with an arbitrary self adjoint realisation
of the differential Laplacian. After discussing spectral properties of 
Laplacians, we prove several versions of trace formulae, relating Laplace 
spectra to sums over periodic orbits on the graph. This includes trace 
formulae with, respectively, absolutely and conditionally convergent periodic 
orbit sums; the convergence depending on properties of the test functions 
used. We also prove a trace formula for the heat kernel and provide 
small-$t$ asymptotics for the trace of the heat kernel.  
\end{abstract}

\newpage

\section{Introduction}
\label{intro}
Some ten years ago Kottos and Smilansky 
\cite{KottosSmilansky:1997,KottosSmilansky:1998} 
introduced quantum graphs as convenient models in the field of quantum chaos
\cite{Haake:1991,Stoeckmann:1999}, where a major goal is to understand the 
connection between dynamical properties of a quantum system and its associated 
classical counterpart \cite{Bohigas:1984}. Previously introduced models that 
possess classical counterparts with chaotic dynamics include quantum billiards, 
motions on Riemannian manifolds with negative curvatures and quantum maps. 
These models have been studied with considerable success, however, they have 
often turned out to bear unwanted complications. Quantum graphs are constructed
along the lines of many of these models in that they are mainly concerned with 
spectral properties of Laplacians. In a sense they are maximally reduced 
versions of such models in that the underlying configuration space is one 
dimensional. The non trivial topology of the graph, however, introduces 
sufficient complexity such that the quantum system behaves like typical 
quantum systems with chaotic classical counterparts, see 
\cite{KottosSmilansky:1997,KottosSmilansky:1998}. On the other hand, many 
details of the classical dynamics are considerably simpler and quantum spectra 
are generally known to some more detail, so that quantum graph models proved 
to be very useful in the field of quantum chaos \cite{Smilansky:2006}.

Trace formulae provide a direct connection between classical and quantum 
dynamics in that they relate quantum spectra to classical periodic orbits.
In general, this connection arises in the form of an asymptotic relation,
valid for large wave numbers. There are only few exceptional cases in which 
trace formulae are identities. Among these are, most notably, Laplacians on 
flat tori and on Riemannian manifolds with constant negative sectional 
curvatures. In both cases the configuration manifolds are Riemannian symmetric 
spaces, allowing for the application of powerful methods of harmonic analysis 
in proving the relevant trace formulae, i.e., the Poisson summation formula and
the Selberg trace formula \cite{Selberg:1956}, respectively.
Generically, however, the tools of harmonic analysis are not available,
and trace formulae have to be proven using semiclassical or microlocal 
techniques, which naturally involve asymptotic methods. Semiclassical trace
formulae were introduced by Gutzwiller \cite{Gutzwiller:1971} for the spectral 
density of quantum Hamiltonians. Subsequently, Balian and Bloch 
\cite{Balian:1972} set up analogous trace formulae for Laplacians on domains 
in $\rz^n$ in a short wavelength approximation. The first mathematical proofs, 
for Laplacians on Riemannian manifolds, are due to Colin de Verdi\`ere 
\cite{Colin:1973} as well as Duistermaat and Guillemin \cite{Duistermaat:1975}.
Later, proofs for the semiclassical case followed 
\cite{Meinrenken:1992,Paul:1995}.

One of the virtues of quantum graph models that led Kottos and Smilansky 
to introduce them to the field of quantum chaos is that their trace formulae
are identities, very much in analogy to the Selberg trace formula. The
first quantum graph trace formula, however, is due to Roth \cite{Roth:1983}, 
who expressed the trace of the heat kernel for a Laplacian with Kirchhoff
boundary conditions in the vertices of the graph as a sum over periodic
orbits. Kottos and Smilansky \cite{KottosSmilansky:1998} then introduced a 
trace formula for the spectral density of the Laplacian, and later Kostrykin, 
Potthoff and Schrader \cite{Schrader:2007} extended Roth's trace formula to 
more general boundary conditions. In these cases the boundary conditions 
characterising the domain of the Laplacian were of a non-Robin type in that 
they do not mix boundary values of functions and their derivatives. This 
leads to periodic orbit sums in the trace formulae that closely resemble 
those occurring in the Selberg trace formula. In this paper our principal 
goal now is to consider general self adjoint realisations of Laplacians on 
compact metric graphs and to prove associated trace formulae with fairly 
general test functions. In particular, we do not require their Fourier 
transforms to be compactly supported. Allowing for Robin-type boundary 
conditions leads to trace formulae that are still identities, yet the 
amplitudes multiplying the oscillating factors in each periodic orbit 
contribution depend on the wave number in a non trivial way. In principle, 
these amplitude functions are known and possess asymptotic expansions for 
large wave numbers. Therefore, in a sense these trace formulae are intermediate
between Selberg and Gutzwiller/Duistermaat-Guillemin trace formulae, where in 
the latter case only the asymptotic expansions of the amplitude functions are 
known, and the test functions must have compactly supported Fourier transforms.

This paper is organised as follows: In section~\ref{prelim} we briefly review
the construction of quantum graphs, including parametrisations of self
adjoint realisations of the Laplacian developed by Kostrykin and 
Schrader~\cite{KostrykinSchrader:1999} as well as Kuchment \cite{Kuchment:2004}.
Following this we investigate various properties of quantum graph edge 
S-matrices as introduced in \cite{KottosSmilansky:1998}, focussing on their 
analytic properties. In section~\ref{gl} we then discuss general properties 
of Laplace spectra on compact graphs. The main part of this paper can be found 
in section~\ref{prw} where we prove several versions of quantum graph trace
formulae: Theorem~\ref{74} contains a trace formula with a double sum over 
periodic orbits, of which only the sum over topological orbit lengths converges 
absolutely, and which allows for a large class of test functions. In 
Theorem~\ref{thm2}, however, we restrict the class of test functions with the 
effect that the entire sum over periodic orbits converges absolutely.
A suitable choice of a test function then allows to establish the trace
of the heat kernel for arbitrary self adjoint realisations of the Laplacian,
see Theorem~\ref{thm3}. We finally summarise and discuss our results in 
section~\ref{concl}. Some of the results presented here have been announced 
in \cite{BolteEndres:2007}.
\section{Preliminaries}
\label{prelim}
We begin with reviewing the relevant concepts underlying the construction
of quantum graphs.
\subsection{Metric graphs}
\label{79}
In the sequel we shall consider finite, metric graphs 
$\Gamma=\left(\mathcal{V},\mathcal{E},\vl\right)$. Here $\mathcal{V}$ is a 
finite set of vertices $\left\{v_1,\ldots,v_V\right\}$ and $\mathcal{E}$ is 
a finite set of edges $\left\{e_1,\ldots,e_E\right\}$. When an edge $e$
connects the vertices $v$ and $w$, these are called edge ends. Two edges are 
adjacent, if they share an edge end; loops and pairs of multiply connected 
vertices shall be allowed. The degree $d_v$ of a vertex $v$ specifies the 
number of edges with $v$ as one of their edge ends. A metric structure can be 
introduced by assigning intervals $[0,l_i]$ to edges $e_i$, along with 
coordinates $x_i\in [0,l_i]$. The $E$-tuple $\vl =(l_1,\dots,l_E)$ then 
collects all edge lengths. Through this procedure the edge ends are mapped 
to the end points of the intervals in a specified manner. The edge end 
corresponding to $x_i =0$ is then called initial point and, correspondingly, 
the other end point is the terminal point of the edge $e_i$. Hence not only 
the connectedness of the graph is specified, but also an orientation of the 
edges is introduced. We emphasise that the specific choice of the orientation 
thus made does not impact the results of this paper, in fact the choice of 
the initial and terminal points of the edges are arbitrary. The intervals are 
only used to construct the Laplace operator as a differential operator.

It is useful to arrange the $2E$ edge ends in a particular way: we list 
the initial points in the order as they occur in the list of edges, followed 
by the terminal points in the same order. For the trace formula we, moreover, 
require the following notions.
\begin{defn}
\label{80}
A \emph{closed path} in $\Gamma$ is a finite sequence of edges 
$\left(e_i\right)_{i=1}^{n}$, such that the edges $e_i$ and $e_{i+1}$ for 
$\in\{1,\ldots,n-1\}$ and the edges $e_n$ and $e_1$ are adjacent. 
A \emph{periodic orbit} is an equivalence class of closed paths modulo cyclic
permutations of the edges. 
\end{defn}
The number $n$ of edges in a periodic orbit $p$ is its \emph{topological 
length}, whereas the sum $l_p = l_{e_1}+\cdots +l_{e_n}$ of the
metric lengths of its edges is the \emph{metric length}, or simply length, 
of $p$. A periodic orbit is primitive, if it is not a multiple repetition 
of another periodic orbit. Furthermore, the set of periodic orbits of the 
graph is called $\cP$, and $\cP_n$ is its subset of orbits with topological 
length $n$.
\subsection{Quantum graphs}
\label{1sec}
Quantum mechanics on a metric graph can be studied in terms of the 
Schr\"odinger equation
\begin{equation}
\label{Schroedinger}
 \ui\frac{\partial}{\partial t}{\psi_t} =-\Delta\psi_t \ .
\end{equation}
Here $\psi_t$ is a vector in an appropriate Hilbert space and
$\Delta$ is a suitable (differential) Laplace operator, acting on 
functions defined on the edges of the graph. Since this operator is of
second order, it is independent of the choice of the coordinates
$x_j$ or $\tilde{x}_j :=l_j-x_j$ on the edges $e_j$, and therefore also 
independent of the specification of edge ends as initial and terminal.

In order to be in a position to view the negative Laplacian in 
\eqref{Schroedinger} as a quantum Hamiltonian, it has to be realised as a
self adjoint operator on a suitable Hilbert space. We therefore introduce
the space of square integrable functions on $\Gamma$ as the quantum graph 
Hilbert space. In this context a function on the graph is a collection 
$F=(f_1,\dots,f_E)$ of functions $f_j : (0,l_j)\to\kz$ on the edges. 
Therefore, one considers the following function spaces,
\begin{equation*}
 C^\infty (\Gamma) = \bigoplus_{j=1}^E C^\infty (0,l_j) \quad\text{and}\quad  
  L^2 (\Gamma) = \bigoplus_{j=1}^E L^2 (0,l_j) \ .
\end{equation*}
The latter is a closed orthogonal sum of Hilbert spaces with respect to the 
scalar product 
\begin{equation*}
\label{def:scalarprod}
 \langle F,G \rangle := \sum_{j=1}^E \int_0^{l_j} \overline{f_j (x_j)}\,
 g_j(x_j)\ \ud x_j \ .
\end{equation*}
As a differential expression the (negative) Laplacian is simply given by
\begin{equation*}
\label{def:diffLapace} 
 -\Delta F := (-f''_1,\dots,-f''_E) \ ,
\end{equation*}
where dashes denote derivatives. This expression may now serve as a way to 
introduce a closed, symmetric operator $(-\Delta,{\mathcal D}_0)$ with domain
\begin{equation*}
\label{def:symdomain}
 {\mathcal D}_0 = \bigoplus_{j=1}^E H^2_0 (0,l_j) \ .
\end{equation*}
Here each term in the orthogonal sum consists of an $L^2$-Sobolev space of 
functions which, together with their derivatives, vanish at the edge ends. 
The deficiency indices of this operator are $(2E,2E)$, and thus it possesses
self adjoint extensions that can be classified according to von Neumann's 
theory (see, e.g., \cite{Reed:1975}). An alternative approach has been 
developed in detail by Kostrykin and Schrader \cite{KostrykinSchrader:1999}, 
which provides a convenient parametrisation that is particularly useful for 
later purposes. In this context one introduces the boundary values 
\begin{equation}
\label{def:bv}
\begin{split}
 F_{bv}  &= \bigl( f_1 (0),\dots,f_E(0),f_1(l_1),\dots,f_E(l_E) \bigr)^T \ ,\\
 F'_{bv} &= \bigl( f'_1 (0),\dots,f'_E(0),-f'_1(l_1),\dots,-f'_E(l_E) 
 \bigr)^T \ ,
\end{split}
\end{equation}
of functions and their derivatives, whereby the signs ensure that inward 
derivatives are considered at all edge ends. Notice that the order of the 
terms follows the convention of arranging edge ends introduced previously. 
Boundary conditions on the functions in the domain of a given self adjoint 
operator are specified through a linear relation between boundary values; 
they are of the form 
\begin{equation}
\label{bc}
 AF_{bv} + BF'_{bv} = 0 \ ,
\end{equation}
see \cite{KostrykinSchrader:1999}.
Here $A,B\in\M(2E,\kz)$ are two matrices such that
\begin{itemize}
\item the matrix $(A,B)$, consisting of the columns of $A$ and $B$, has 
maximal rank $2E$,
\item $AB^{\ast}$ is self adjoint.
\end{itemize}
These conditions then imply the self adjointness of the operator, and every 
self adjoint extension can be achieved in this manner. Occasionally, we shall 
denote a particular such self adjoint realisation of the Laplacian as 
$\Delta(A,B,\vl)$. This parametrisation, however, is obviously not unique
because a multiplication of (\ref{bc}) with $C\in\GL(2E,\kz)$ from the left 
does not change the boundary conditions. On the other hand, if 
$\Delta(A,B,\vl)=\Delta(A',B',\vl)$, there exists $C\in\GL(2E,\kz)$ with 
$A'=CA$ and $B'=CB$, see \cite{KostrykinSchrader:1999}. Thus, for any 
$C\in\GL(2E,\kz)$ both $A,B$ and $A'=CA,B'=CB$ provide an equivalent 
characterisation of the same operator.

The linear relations (\ref{bc}) can in principle relate boundary values at 
any set of edge ends. We wish, however, the operator to respect the 
connectedness of the graph and therefore we restrict ourselves to local 
boundary conditions. These are characterised by the condition that
\eqref{bc} only relates edge ends that form a single vertex. To this end we 
now group the edge ends in (\ref{def:bv}) according to the vertices they belong 
to. Local boundary conditions then lead to a block structure of the matrices 
$A$ and $B$,
\begin{equation}
\label{localbc}
 A = \bigoplus_{v\in{\mathcal V}}A_v \quad\mathrm{and}
 \quad B = \bigoplus_{v\in{\mathcal V}}B_v \ ,
\end{equation}
such that each block, represented by $A_v$ and $B_v$, exactly relates the 
boundary values of functions and their derivatives at the vertex $v$. In 
this context self adjointness of the Laplacian is achieved, if for all 
$v\in {\mathcal V}$ the rank of $(A_v,B_v)$ is $d_v$ and $A_v B_v^*$ is self 
adjoint. 

To mention a few examples, a vertex with Dirichlet boundary conditions
can be characterised by $A_v = \eins_{d_v}$, $B_v =0$, whereas for Neumann
boundary conditions one would choose $A_v =0$, $B_v = \eins_{d_v}$. 
Moreover, the generalised Kirchhoff boundary conditions used by Kottos and 
Smilansky \cite{KottosSmilansky:1998} can be achieved by choosing
\begin{equation}
\label{KirchAB}
 A_v = \begin{pmatrix} 1 & -1 & & \\  & \ddots & \ddots & \\ & & 1 & -1\\ 
                       & & & \mu_v \end{pmatrix}\quad\text{and}\quad 
 B_v = \begin{pmatrix}  & & & \\  &  &  & \\ & & & \\ 
                        1 & \cdots & \cdots & 1 \end{pmatrix} \ ,
\end{equation}
where only the non vanishing matrix entries are indicated. Here $\mu_v$ must 
be real; when $\mu_v =0$ the usual Kirchhoff conditions are realised.

The approach we have taken here when singling out local boundary conditions 
is to start from a given graph and then to realise the Laplacian as a self
adjoint operator on that graph. An alternative view would be to consider the 
above construction as a realisation of the Laplacian as a self adjoint operator 
on a collection of $E$ intervals. One can then find a graph, i.e., a way to 
connect the interval ends in vertices, such that a given pair of matrices 
$A$ and $B$ is local with respect to this graph; there even is a unique graph
maximising the number of vertices \cite{KostrykinSchrader:2006}. In the 
following we shall, however, stick to the previous view.

The non-uniqueness in the choice of the matrices $A$ and $B$ can be overcome 
by parametrising the self adjoint realisations of the Laplacian in terms of 
projectors onto subspaces of the $2E$-dimensional spaces of boundary values.  
To this end Kuchment \cite{Kuchment:2004} introduced the projector $P$ onto 
the kernel of $B$ as well as the projector $Q=\eins-P$ onto the orthogonal
complement $(\ker B)^\perp =\ran B^\ast$ in $\kz^{2E}$ and proved that $A$
maps $\ran B^\ast$ into $\ran B$. He then defined the (self adjoint) 
endomorphism
\begin{equation}
\label{def:L}
L := \bigl( B|_{\ran B^\ast} \bigr)^{-1} AQ
\end{equation}
of $\ran B^\ast$, and showed that the boundary conditions (\ref{bc}) are 
equivalent to
\begin{equation}
\label{altbc}
 PF_{bv} = 0 \qquad\text{and}\qquad LQF_{bv} + QF'_{bv} = 0 \ .
\end{equation}
Moreover, there exists a $C\in\GL(2E,\kz)$ such that 
\begin{equation}
\label{1}
A'=CA=P+L \qquad \text{and} \qquad B'=CB=Q, 
\end{equation}
implying that 
\begin{equation*}
\label{63}
 L=A' B'^\ast.
\end{equation*}
A refinement of this construction can be found
in \cite{Fulling:2007}. From (\ref{altbc}) one concludes that in cases where 
$L=0$, the boundary conditions do not mix boundary values of the functions 
themselves with those of their derivatives. We call these 
{\it non-Robin boundary conditions}, and all other cases {\it Robin boundary 
conditions}.  

Yet another description of the self adjoint realisations of the Laplacian
employs the associated quadratic forms. As shown by Kuchment 
\cite{Kuchment:2004}, given a realisation of the Laplacian as a self adjoint 
operator, the associated quadratic form 
\begin{equation}
\label{Qform}
 Q_\Delta [F] = \sum_{e=1}^E \int_0^{l_e}| f_e'(x)|^2 \ \ud x - 
 \langle F_{bv},L F_{bv}\rangle_{\kz^{2E}} 
\end{equation}
has a domain that consist of all functions $F$ on the graph, with components 
$f_e\in H^1(0,l_e)$, whose boundary values fulfil $PF_{bv}=0$.
\section{The S-matrix}
\label{2sec}
In quantum graph models Laplace eigenvalues can be conveniently characterised 
in terms of zeros of finite dimensional determinants, and thus these models 
are amenable to powerful analytical as well as numerical methods. In quantum 
billiards a related method was pioneered by Doron and Smilansky~\cite{DorSmi92}
as the scattering approach to quantisation. In general, this method relies 
on semiclassical approximations. As first demonstrated by Kottos and 
Smilansky~\cite{KottosSmilansky:1998}, however, in quantum graphs the 
scattering approach allows to determine Laplace eigenvalues exactly from a 
finite dimensional secular equation.

The scattering approach bears its name from the fact that it is based on
scattering processes occurring when one opens up a given closed quantum
system appropriately. In quantum graphs the procedure of opening up
consists of replacing each vertex and its attached edges by an infinite 
star graph. This is the single, given vertex $v$ with $d_v$ infinite half
lines attached that replace the edges of finite lengths. Carrying over the
local boundary conditions at the vertices from the original closed quantum 
graph, one thus obtains $V$ open quantum systems, which each possess an 
on-shell scattering matrix $\sigma^v(k)$. More precisely, this S-matrix is 
defined in terms of $d_v$ functions $F^{(j)}$ on the infinite star graph 
associated with $v$, whose components on the $d_v$ infinite edges are
\begin{equation*}
 f^{(j)}_{i}(x):= 
 \begin{cases} \sigma^v_{ji}(k)\,\ue^{\ui kx} \ , & j\neq i \\
 \ue^{-\ui kx} + \sigma^v_{jj}(k)\,\ue^{\ui kx} \ , & j=i
 \end{cases} \ .
\end{equation*}
One then requires each of these functions to fulfil the boundary conditions at 
$v$. In terms of the parameterisation 
of the boundary conditions described in Section~\ref{1sec} one then finds 
that
\begin{equation}
\label{vertexSmatrix}
 \sigma^v (k) = -(A_v + \ui k B_v )^{-1} (A_v - \ui k B_v) \ , \quad\text{for}
 \quad k\in\rz\backslash\{0\} \ .
\end{equation}
The conditions imposed on $A_v$ and $B_v$ in order to achieve self adjoint
boundary conditions ensure that $A_v\pm\ui k B_v$ are invertible and that the 
vertex S-matrix $\sigma^v (k)$ is unitary for all $k\in\rz\setminus\{0\}$,
see \cite{KostrykinSchrader:1999}. 

The local scattering matrix associated with a vertex with Dirichlet or
Neumann boundary conditions is $\sigma^v=-\eins_{d_v}$ or 
$\sigma^v=\eins_{d_v}$, respectively. In contrast, according to (\ref{KirchAB})
generalised Kirchhoff boundary conditions lead to a vertex S-matrix with
elements of the form \cite{KottosSmilansky:1998}
\begin{equation}
\label{genKirchSm}
 \sigma^v_{ij} (k) =  -\delta_{ij} + \frac{2k}{d_v k+\ui\mu_v}\ .
\end{equation}

The local S-matrices of the entire graph can now be grouped together vertex 
by vertex. In that process all edges occur twice, namely associated with the 
vertex S-matrices of their two edge ends. It is hence useful to consider
directed edges, and view the matrix elements of $\sigma^v$ as describing
transitions from a directed edge with terminal point $v$ to a directed edge 
that has $v$ as its initial point. As the list of all directed edges  
corresponds to listing their initial points, the transitions prescribed
by $\sigma^v$ can also be performed on the boundary values \eqref{def:bv}.
As a result one obtains the matrix
\begin{equation}
\label{def:Smatrix}
 S(A,B;k) = -(A+\ui k B)^{-1}(A-\ui kB) \ .
\end{equation}
Again, $A\pm\ui k B$ are invertible and $S(A,B;k)$ is unitary for all 
$k\in\rz\setminus\{0\}$. Moreover, \eqref{def:Smatrix} is invariant under
the substitution of $A,B$ by $CA,CB$ for all $C\in\GL (2E,\kz)$ and 
therefore is associated with the self adjoint realisations of the Laplacian
\cite{KostrykinSchrader:2006}.

We emphasise that, although \eqref{def:Smatrix} can be defined for any self 
adjoint realisation, the vertex S-matrices \eqref{vertexSmatrix} can only be 
recovered from \eqref{def:Smatrix} in the case of local boundary conditions, 
i.e., when \eqref{localbc} is satisfied. Despite the fact that a closed 
quantum graph does not allow for quantum scattering in the usual sense, the 
quantity (\ref{def:Smatrix}) is often referred to as the edge (or bond) 
S-matrix \cite{KottosSmilansky:1998} of the quantum graph.

Furthermore, using the parametrisation \eqref{altbc} of boundary conditions 
and utilising \eqref{1} as well as the fact that $(L+\ui k)^{-1}$ commutes 
with $L-\ui k$, one obtains the representation
\begin{equation}
\label{SthruPQ}
S(A,B;k) = -P - Q\, (L+\ui k)^{-1}(L-\ui k) \, Q \ , 
          \quad k\in\rz\backslash\{0\} \ ,
\end{equation}
for the edge S-matrix, see \cite{KostrykinSchrader:2006}.

From the expressions \eqref{def:Smatrix} and \eqref{SthruPQ} it appears that 
the S-matrix generally depends on the wave number $k$ in a non-trivial way. 
However, certain boundary conditions lead to $k$-independent S-matrices. 
Obvious examples are Dirichlet and Neumann boundary conditions, as well as the
usual Kirchhoff boundary conditions, i.e., \eqref{genKirchSm} with $\mu_v =0$. 
A general characterisation of such boundary conditions was provided
in \cite{Schrader:2007} in terms of the following equivalent conditions:
\begin{itemize}
\item $S(A,B;k)$ is $k$-independent.
\item $S(A,B;k)$ is self adjoint for some, and hence for all, $k>0$.
\item $S(A,B;k)=\eins -2P$ for some, and hence for all, $k>0$.
\item $AB^\ast =0$, i.e., $L=0$.
\end{itemize}
The last point shows that $k$-independent S-matrices arise exactly in the case
of non-Robin boundary conditions.

Below we are going to prove some properties of S-matrices that are relevant 
for the trace formula. In this context, for Robin boundary conditions an 
important role will be played by the spectrum $\sigma(L)$ of the self adjoint 
matrix L \eqref{def:L}. 

We shall make extensive use of the S-matrix extended to complex wave numbers 
$k$ and therefore need the following.
\begin{lemma}
\label{3}
Let $A$ and $B$ specify self adjoint boundary conditions for the Laplacian 
on the graph. Then the S-matrix (\ref{def:Smatrix}) has the following 
properties:
\begin{enumerate}
\item $S(A,B;k)$ can be continued into the complex $k$-plane as a meromorphic 
function, and has simple poles at the points of the set 
$\ui\sigma(L)\setminus\{0\}$.
\item $S(A,B;k)$ is unitary for all $k\in\rz$.
\item $S(A,B;k)$ is invertible for all $k\in\sm$, and its inverse is 
$S(A,B;-k)$.	
\end{enumerate}
\end{lemma}
\begin{proof}
We henceforth extend the self adjoint endomorphism $L$ of $\ran B^\ast$,
see \eqref{def:L}, to an endomorphism of $\kz^{2E}$ by setting it to zero
on $(\ran B^\ast)^\perp =\ker B$. We then diagonalise $L$ utilising an 
appropriate unitary $W$, and denote the non-zero eigenvalues (counted with
their multiplicities) by $\{\lambda_1,\ldots,\lambda_d\}$. This leaves the 
eigenvalue zero with a multiplicity of $2E-d$. There are 
$r:=\dim\ran B^\ast -d$ (orthonormal) eigenvectors of $L$ in $\ran B^\ast$ 
and $s:=\dim\ker B=2E-\dim\ran B^\ast$ eigenvectors in $\ker B$, respectively, 
corresponding to the eigenvalue zero.

Employing this diagonalisation in the representation \eqref{SthruPQ} of the 
S-matrix then leads to the expression 
\begin{equation}
\label{5a}
  S(A,B;k)=W^{\ast}
	   \begin{pmatrix}
	     \begin{matrix}
	     -\frac{\lambda_1-\ui k}{\lambda_1+\ui k} & & \\
	       & \ddots & \\
	       & & -\frac{\lambda_d-\ui k}{\lambda_d+\ui k}
	     \end{matrix}
	    & & \\ & \eins_r & \\ & & -\eins_s
	   \end{pmatrix}
	   W \ .
\end{equation}
Since the unitary $W$ is independent of $k$ the first statement of the lemma
is obvious.

The unitarity of $S(A,B;k)$ for real $k$ also follows immediately by observing 
that the diagonal entries in \eqref{5a} are all of unit absolute value.

The third statement follows in a completely analogous fashion from the
representation \eqref{5a}.
\end{proof}
Knowing that the S-matrix is analytic in $k$, one would like to calculate
its derivative. This in fact is required in the proof of the trace formula
below. It is even possible to relate the derivative of $S(k)$ to the
S-matrix itself.
\begin{lemma}
\label{16}
Under the same assumptions as in Lemma~\ref{3} one obtains for
$k\in\sm$,
\begin{equation}
\label{Sderiva}
 \dk S(A,B;k) = -\frac{1}{2k}\left[S(A,B;k)-S(A,B,k)^{-1}\right]S(A,B;k) \ .
\end{equation}
\end{lemma}
We remark that for real $k$ the unitarity of the S-matrix can be invoked 
to obtain from \eqref{Sderiva} that it is independent of $k$, iff it is self 
adjoint.
\begin{proof}
Let us first assume that $k\in\rz\setminus\{0\}$ and abbreviate $S(A,B;k)$ 
as $S(k)$. We also denote derivatives w.r.t. $k$ by a dash and use the 
relation $\dk [X(k)]^{-1} = -X(k)^{-1}X'(k) X(k)^{-1}$, which is true
for any differentiable function $X(k)$ taking values in $\GL(2E,\kz)$.
Recall that the conditions imposed on $A,B$ ensure that $A\pm\ui kB$ is 
invertible for $k\in\rz\setminus\{0\}$. Hence
\begin{equation}
\label{Sderi}
\begin{split}
 S'(k) &= (A+\ui k B)^{-1}(\ui B)(A+\ui k B)^{-1}(A-\ui k B) +
          (A+\ui k B)^{-1}(\ui B) \\
       &= -\ui (A+\ui k B)^{-1}B\bigl( S(k)-\eins \bigr) \ .
\end{split}
\end{equation}
The last line is obviously invariant under a replacement of $A$ and $B$
by $CA$ and $CB$, respectively, where $C\in\GL(2E,\kz)$ is arbitrary.
  
Now choose $C(k)=(A+\ui k B)^{-1}\in\GL(2E,\kz)$ and find that 
(see also \cite{KostrykinSchrader:2006b})
\begin{equation}
\label{ABprime}
 C(k)A = -\frac{1}{2}\bigl( S(k)-\eins \bigr) \quad\text{and}\quad
 C(k)B = -\frac{1}{2\ui k} \bigl( S(k)+\eins \bigr) \ .
\end{equation}
Inserting this into (\ref{Sderi}) finally leads to
\begin{equation}
\label{SderivR}
\begin{split}
 S'(k) &= \frac{1}{2k} \bigl(  S(k)+\eins \bigr)\bigl( S(k)-\eins \bigr) \\
       &= -\frac{1}{2k} \bigl( S(k) - S(k)^\ast \bigr) \, S(k) \ ,
\end{split}
\end{equation}
which proves the statement for $k\in\rz\setminus\{0\}$.

From Lemma~\ref{3} we infer that $S(k)$ is analytic in a neighbourhood of
$k=0$ and that the right-hand side of \eqref{SderivR} has a removable 
singularity at $k=0$; hence \eqref{SderivR} extends to all real $k$.
Since, moreover, $S(k)$ is unitary on $\rz$ and analytic on $\smz$, and 
$\frac{1}{k}$ as well as $S(k)^{-1}$ are also analytic on this set, the full
statement of the lemma follows by analytic continuation.
\end{proof}
From Lemma~\ref{3} we know that $S(k)$ is meromorphic with finitely many 
poles on the imaginary axis. One can therefore perform power series 
expansions of the S-matrix in large parts of the complex $k$-plane. Below
we want to specify two such expansions, and to this end we exclude an annulus
containing $\ui\sigma(L)\setminus\{0\}$. This annulus is characterised by
the two radii 
\begin{equation*}
 \lambda_{\min/\max}:=\min/\max\{|\lambda|;\ \lambda\in
 \sigma(L)\setminus\{0\}\} \ . 
\end{equation*}
Here we assume that $\sigma(L)\neq\{0\}$; otherwise $L=0$ which is equivalent 
to the S-matrix being independent of $k$. 

We are now in a position to provide the announced expansions of the S-matrix.
\begin{lemma}
\label{70}
Let the same conditions as in Lemma~\ref{3} be given and assume that
$L\neq 0$. Then the following power series expansions converge absolutely and 
uniformly in any closed subsets of the specified regions:
\begin{enumerate}
\item For $|k|>\lambda_{\max}$,
\begin{equation}
\label{expan1}
 S(A,B;k)=\eins-2P+2\sum_{n=1}^\infty \frac{1}{k^n}\,(\ui L)^n \ .  
\end{equation}
\item For $|k|<\lambda_{\min}$,
\begin{equation}
\label{expan2}
 S(A,B;k)=-\eins+2\tilde{P}-2\sum_{n=1}^\infty k^n \,
          \left(\ui \tilde L \right)^n\ ,
\end{equation}
where $\tilde P$ and $\tilde L$ emerge from $P$ and $L$, respectively, by
replacing $A,B$ with $-B,A$.
\end{enumerate}
\end{lemma}
\begin{proof}
For the first expansion we refer to the representation \eqref{5a} of the
S-matrix and employ the expansion
\begin{equation*}
\label{Leigenv}
 -\frac{\lambda_\alpha -\ui k}{\lambda_\alpha +\ui k} = 
 \frac{1+\frac{\ui}{k}\lambda_\alpha}{1-\frac{\ui}{k}\lambda_\alpha} =
 1 + 2 \sum_{n=1}^\infty \Bigl(\frac{\ui\lambda_\alpha}{k}\Bigr)^n \ , 
 \quad \alpha=1,\dots,d \ ,
\end{equation*}
valid for $|k|>\lambda_\alpha$. Hence, for $|k|>\lambda_{\max}$ the S-matrix 
is $\eins +2\sum_{n=1}^\infty \frac{1}{k^n}\,(\ui L)^n$ on $\ran B^\ast$ and 
$-\eins$ on $\ker B$. Since $L=0$ on $\ker B$, the relation~\eqref{expan1} 
follows.

For the second expansion we remark that from \eqref{def:Smatrix} one can 
readily deduce the relation $S(A,B;k)=-S(-B,A;\tfrac{1}{k})$ for 
$k\in\rz\setminus\{0\}$, see~\cite{KostrykinSchrader:2006}. Thus, 
\eqref{expan1} implies \eqref{expan2}, and the domain $|k|<\lambda_{\min}$ 
of convergence, in which $S(k)$ is holomorphic, follows immediately.
\end{proof}
Lemma~\ref{70} also provides limiting expressions for the edge S-matrix as
$|k|\to\infty$ and $|k|\to 0$, respectively,
\begin{equation*}
 S_\infty = \eins -2P \quad\text{and}\quad S_0 = -\eins + 2\tilde P \ ,
\end{equation*}
which we shall use subsequently.

Later we shall integrate expressions containing the S-matrix along contours
in the upper complex half plane and, therefore, we need to estimate the norm of
the S-matrix along the contours. To this end we introduce
\begin{equation}
\label{lambdaminplus}
 \lambda_{\min}^+ := 
 \begin{cases} \min\{\lambda\in\sigma(L);\ \lambda>0\}\ , & \text{if} \  \exists
                           \lambda_\alpha >0\\
                           \infty\ , & \text{else} \end{cases} \ ,
\end{equation}
and obtain the following.
\begin{lemma}
\label{lem:Sextend}
Let $k\in\rz$ and $0<\kappa<\lambda_{\min}^+$, then
\begin{equation}
\label{Sbound2}
 \| S(k+\ui\kappa) \| , \  \| S(k-\ui\kappa)^{-1} \|
 \leq \max \left\{ 1,
 \frac{\lambda^+_{\min}+\kappa}{\lambda^+_{\min}-\kappa}
 \right\}
\end{equation}
in the operator norm. Furthermore, if $\kappa>\lambda_{\max}$, then
\begin{equation}
\label{Sbound3}
 \| S(k+\ui\kappa) \| , \  \| S(k-\ui\kappa)^{-1} \|
 \leq \frac{\kappa +\lambda_{\max}}{\kappa-\lambda_{\max}} \ .
\end{equation}
\end{lemma}
\begin{proof}
From \eqref{5a} one can read off the eigenvalues of $S(k+\ui\kappa)$ as 
$\pm 1$ and $-(\lambda_\alpha+\kappa-\ui k)/(\lambda_\alpha-\kappa+\ui k)$, 
$\alpha=1,\dots,d$.  In absolute values the latter quantities, as functions 
of $k\in\rz$, are maximised at $k=0$. Now suppose that $\lambda^+_{\min}>0$ 
and let $0<\kappa<\lambda^+_{min}$. Then the largest quantity among the 
$|\lambda_\alpha +\kappa|/|\lambda_\alpha -\kappa|$ is the one on the 
right-hand side of (\ref{Sbound2}). In the case 
$\lambda^+_{\mathrm{min}}=\infty$ the upper bound is one. The proof for 
$S(k-\ui\kappa)^{-1}$ is completely analogous.

If $\kappa>\lambda_{\max}$ the same argument yields the bound \eqref{Sbound3}.
\end{proof}
\section{The spectrum of the Laplacian}
\label{gl}
The scattering approach to the quantisation of a finite, metric graph 
utilises a secular equation based on the edge S-matrix of the graph. 
Here we closely follow the original approach as developed by Kottos 
and Smilansky \cite{KottosSmilansky:1998} for the case of (generalised) 
Kirchhoff boundary conditions, which was later generalised  by 
Kostrykin and Schrader \cite{KostrykinSchrader:2006b}. To keep
the presentation sufficiently self-contained, we reproduce the relevant
results below. We begin, however, with some general properties of Laplace 
spectra and finish this section with some remarks on the eigenvalue zero.
\subsection{Preliminaries on the spectrum}
Given a Laplacian on a compact metric graph, one would naturally expect
that its spectrum is discrete and has a finite lowest eigenvalue. 
Kuchment indeed proved \cite{Kuchment:2004} that any such self adjoint 
(negative) Laplacian is bounded from below, and that its resolvent is trace 
class. Thus, the spectrum is discrete and bounded from below. Subsequently,
Kostrykin and Schrader \cite{KostrykinSchrader:2006b} improved the lower
bound. They showed that
\begin{equation}
\label{specbound}
 -\Delta\geq -s^2 \ ,
\end{equation}
where $s\geq 0$ is the unique solution of
\begin{equation}
\label{boundcond}
 s\tanh\left(\frac{s l_{\min}}{2}\right)  = \lambda_{\max}^+ \ ,
\end{equation}
with $l_{\min}$ denoting the shortest edge length and
\begin{equation}
\label{lmp+def}
 \lambda_{\max}^+ := 
        \begin{cases} 
        \max\{\lambda\in\sigma(L);\ \lambda>0\}\ , & \text{if} \  \exists
        \lambda_\alpha >0 \\ 0 , & \text{else} 
        \end{cases} \ .
\end{equation}
\begin{rem}
As an aside we should like to mention that the lower bound \eqref{specbound} 
is optimal in the following sense: Consider a trivial example of a metric graph
given by an interval $I$ of length $l$, and a Laplacian with domain specified by
the choice
\begin{equation*}
 A=\lambda\eins_2 \quad\text{and}\quad B =\eins_2 \ ,
\end{equation*}
where $\lambda>0$, such that $L=\lambda\eins_2$ and $\sigma(L)=\{\lambda\}$. 
(Equivalently, $P=0$, $Q=\eins_2$ and $L=\lambda\eins_2$.) Hence,
\eqref{altbc} leads to the Robin boundary conditions
\begin{equation*}
 \lambda \begin{pmatrix} f(0) \\ f(l) \end{pmatrix} +
 \begin{pmatrix} f'(0) \\ -f'(l) \end{pmatrix} = 0 \ ,
\end{equation*}
and this implies the quantisation condition
\begin{equation}
\label{Robinquant}
 \left( \frac{\lambda -\ui k}{\lambda + \ui k} \right)^2\, 
 \ue^{2\ui kl} = 1 \ .
\end{equation}
The solution $k=\ui\kappa$, with $\kappa>\lambda>0$, representing the lowest 
Laplace eigenvalue $-\kappa^2$ corresponds to a solution of the equation 
\begin{equation*}
 \kappa\tanh\left(\frac{\kappa l}{2}\right)  = \lambda \ .
\end{equation*}
This condition is equivalent to \eqref{boundcond}, demonstrating that the bound
\eqref{specbound} is sharp for this `quantum graph'. 
\end{rem}
Kostrykin and Schrader also showed \cite{KostrykinSchrader:2006b} that the 
number of negative Laplace eigenvalues is bounded by the number of positive 
eigenvalues of $L$ (counted with their respective multiplicities). In the 
example above, the Robin Laplacian on an interval hence has at least one and 
at most two negative eigenvalues.

For the trace formula one requires an a priori estimate on the number of 
eigenvalues. This is well known for the Dirichlet and the Neumann Laplacian,
and in the case of Kirchhoff boundary conditions can be found in
\cite{Solomyak:2002}. The same asymptotic law, however, holds also in the 
general case.
\begin{prop}
\label{Weylnew}
Given a self adjoint realisation of the Laplacian on a compact metric graph,
the number of its eigenvalues $k_j^2\in\rz$ (counted with their multiplicities)
fulfils the following asymptotic law,
\begin{equation}
\label{Weyllawnew}
 N(K):= \# \left\{ j;\ k_j^2 \leq K^2 \right\} \sim 
        \frac{\cL}{\pi}\,K \ , \quad K\to\infty \ ,
\end{equation}
where $\cL := l_{e_1}+\cdots + l_{e_E}$ is the total length of the graph.
In particular, the Laplacian has infinitely many eigenvalues that only
accumulate at infinity.
\end{prop}
\begin{proof}
We prove the asymptotic law employing a variational characterisation of
the eigenvalues based on the quadratic form~\eqref{Qform}, as well as an 
analogue of the Dirichlet-Neumann bracketing (see \cite{Reed:1978}). To
this end we introduce two comparison operators. 

The first comparison operator is the direct sum of the Dirichlet operators 
on the edges, i.e., the Dirichlet-Laplacian of section~\ref{1sec}. The domain 
of the associated quadratic form is characterised by the condition $F_{bv}=0$, 
and therefore is contained in the domain of \eqref{Qform}. Moreover, on the
Dirichlet-form domain both quadratic forms coincide. The comparison lemma
devised in \cite{Reed:1978} hence implies that $N_D(K)\leq N(K)$. Here 
$N_D(K)$ is the counting function for the eigenvalues of the 
Dirichlet-Laplacian, which trivially fulfils the asymptotic law 
\eqref{Weyllawnew}.

As our second comparison operator we choose the direct sum of Robin Laplacians
on the edges (see the example above). This operator is characterised by 
$P_R=0$, $Q_R=\eins_{2E}$ and $L=\lambda\eins_{2E}$ with some $\lambda\geq 0$. 
($\lambda=0$ in fact corresponds to Neumann Laplacians on the edges.) 
The boundary conditions are $\lambda F_{bv}+F'_{bv}=0$, and thus they decouple 
the edges. The respective eigenvalue counting function $N_R(K)$ can be 
determined from \eqref{Robinquant} and clearly obeys the asymptotic law 
\eqref{Weyllawnew}. The associated form domain, characterised by the condition 
$P_R F_{bv}=0$, contains the domain of \eqref{Qform}. Choosing 
$\lambda=\lambda_{\max}^+$, see \eqref{lmp+def}, on the form domain of 
\eqref{Qform} one finds $Q_R[F]\leq Q_\Delta[F]$. Therefore, the comparison 
lemma of \cite{Reed:1978} implies $N_R(K)\geq N(K)$.

Thus, $N_D(K)\leq N(K)\leq N_R(K)$ and the upper and the lower bounds both
fulfil the same asymptotic law, which proves \eqref{Weyllawnew}. The further
statement follows immediately.
\end{proof}
This result is independent of any details of the quantum graph, apart from 
its volume. This is analogous to the corresponding results for Laplacians
on manifolds or domains. In general, the asymptotic growth of the number of
eigenvalues is proportional to the volume of the manifold/domain and to
$K^D$, where $D$ is the dimension of the manifold/domain. This type of
results is often referred to as `Weyl's law' \cite{Weyl:1911}, and insofar 
Proposition~\ref{Weylnew} is the quantum graph version of Weyl's law. 
\subsection{The secular equation}
Apart from the edge S-matrix, the scattering approach requires the metric 
information of the graph, which enters through
\begin{equation}
\label{def:Tofk}
 T(\vl;k) := 
 \begin{pmatrix} 0 & t(\vl;k) \\ t(\vl;k) & 0 \end{pmatrix} 
 \quad\text{with} \quad t(\vl;k) := 
 \begin{pmatrix} \ue^{\ui kl_1} & & \\ & \ddots & \\ & & \ue^{\ui kl_E}
 \end{pmatrix} \ ,
\end{equation}
where $k\in\kz$. Both matrices are then used to introduce
\begin{equation}
\label{def:Uofk}
 U(k) := S(A,B;k)\,T(\vl;k) \ .
\end{equation}
The topological and the metric data entering $U(k)$ are hence clearly 
separated.  

For real $k$ the endomorphisms $S(k)$, $T(k)$ and $U(k)$ of $\kz^{2E}$ 
are obviously unitary. We therefore denote the eigenvalues of $U(k)$ by  
$\ue^{\ui\theta_1(k)},\dots,\ue^{\ui\theta_{2E}(k)}$. Following Lemma~\ref{3} we 
conclude that $U(k)$ can be extended into the complex $k$-plane as a 
meromorphic function with poles at $\ui\sigma(L)\setminus\{0\}$. The 
determinant function
\begin{equation}
\label{23}
 F(k):=\det\bigl(\eins-U(k)\bigr) \ ,
\end{equation}
on which the scattering approach is based (see \cite{KottosSmilansky:1998}),
hence is also meromorphic. Its poles are in $\ui\sigma(L)\setminus\{0\}$,
but do not necessarily exhaust the entire set.
\begin{prop}[Kostrykin, Schrader \cite{KostrykinSchrader:2006b}]
\label{thm:seceq}
The determinant function \eqref{23} is meromorphic on the complex plane with 
poles in the set $\ui\sigma(L)\setminus\{0\}$. Furthermore, let $k_n\in\smu$ 
with $\im k_n\geq 0$, then $k_n^2$ is an eigenvalue of $-\Delta$, iff $k_n$ 
is a zero of the function~\eqref{23}, i.e., $F(k_n)=0$. Moreover, the spectral 
multiplicity  $g_n$ of the Laplace eigenvalue $k_n^2 >0$ coincides with the 
multiplicity of the eigenvalue one of $U(k_n)$.
\end{prop}

Proposition~\ref{thm:seceq} establishes a close connection between zeros of 
the determinant function \eqref{23} and Laplace eigenvalues. Notice that 
although Laplace eigenvalues occur as squares, $k^2$, the function \eqref{23} 
is not invariant under a change of sign in its argument. There exists, however,
a functional equation under the substitution $k\mapsto -k$.
\begin{lemma}
\label{functeq}
For all $\sm$ the following identity holds:
\begin{equation}
\label{funceq}
 F(k) = (-1)^M \, \ue^{2\ui k\cL} \left( \prod_{\alpha=1}^d
 \frac{\lambda_\alpha -\ui k}{\lambda_\alpha +\ui k}\right) F(-k) \ ,
\end{equation}
where $M=E+d+\dim\ker B$ and $\cL=l_{e_1}+\cdots +l_{e_E}$ is the sum of 
all edge lengths.
\end{lemma}
\begin{proof}
We decompose
\begin{equation*}
 \eins - S(k)T(k) = - S(k) \bigl[ \eins - S(-k)T(-k) \bigr] T(k) \ ,
\end{equation*}
so that after taking determinants,
\begin{equation*}
 F(k) = (-1)^{2E}\det T(k) \, \det S(k) \, F(-k) \ .
\end{equation*}
Using the definition \eqref{def:Tofk} of $T(k)$ and the representation
\eqref{5a} of $S(k)$ then yields \eqref{funceq}.
\end{proof}
We remark that when $k^2\in\rz$ is a Laplace eigenvalue, then either $k\in\rz$
when the eigenvalue is non-negative or, in the case of a negative eigenvalue,
$k=\pm\ui\kappa$ with $\kappa>0$. The relevant zeros of the determinant
\eqref{23} are then $\pm k\in\rz$ or $\ui\kappa$, respectively. Unless 
$\kappa\in\sigma(L)$ or $k= 0$, all Laplace eigenvalues are covered by 
Proposition~\ref{thm:seceq}. However, in order to count Laplace eigenvalues
in terms of zeros of $F(k)$ with their correct multiplicities one has to
establish a connection between the order of the zero and the multiplicity of 
one as an eigenvalue of $U(k)$. In the trace formula we shall need this
connection for the non-negative eigenvalues and hence now consider the 
eigenphases $\theta(k)$ of $U(k)$, defined through
\begin{equation}
\label{15}
 U(k) v(k)=\ue^{\ui\theta(k)}v(k) \quad \text{with} \quad \|v(k)\|=1 \ ,
\end{equation}
for $k\in\rz$. We then recall that $U(k)=S(A,B;k)T(\vl;k)$ is analytic in 
$\kz\setminus[\ui\sigma(L)\setminus\{0\}]$. According to analytic perturbation 
theory (see, e.g., \cite{Kato:1980}) its eigenvalues, for which we keep the 
notation $\ue^{\ui\theta(k)}$, are continuous on this set and differentiable 
apart from possibly isolated points. Since, however, $U(k)$ is real-analytic
and normal for all $k\in\rz$, we can apply a sharpened version of analytic
perturbation theory (see \cite{Lancaster:1985}) to conclude that the 
eigenvalues are real analytic for all $k\in\rz$ and that there exists a choice 
of eigenvectors with the same property. 

The following statement is a generalisation of a result 
found in \cite{KottosSmilansky:1998,Berkolaiko:2008b} that is valid for 
$k$-independent S-matrices.
\begin{lemma}
\label{17}
Let $\theta(k)$ be an eigenphase of $U(k)$, $k\in\rz$, with associated 
normalised eigenvector $v(k)=(v_1(k),\dots,v_{2E}(k))^T$. Then
\begin{equation}
\label{kderivlem}
 \dk\theta(k) = \sum_{i=1}^{2E} l_i |v_i(k)|^2 - 
                 2\left<v(k),\frac{L}{L^2+k^2}v(k)\right>_{\kz^{2E}} \ .
\end{equation}
\end{lemma}
\begin{proof}
Taking Lemma \ref{16} into account we first observe that
\begin{equation}
\label{Udash}
\begin{split}
 U'(k) &= -\frac{1}{2k}\bigl( S(k)-S^\ast (k)\bigr) S(k)T(k)
          +\ui S(k)T(k)D(\vl) \\
       &= -2\ui \frac{L}{L^2 +k^2} U(k) + \ui U(k)D(\vl) \ , 
\end{split}
\end{equation}
where
\begin{equation}
\label{diagL}
 D(\vl) := \begin{pmatrix} D_1(\vl) & 0 \\ 0 & D_1(\vl) \end{pmatrix} \ ,
 \qquad  D_1(\vl) := 
 \begin{pmatrix} l_1 & & \\ &\ddots& \\ & & l_E \end{pmatrix} \ .
\end{equation}
We also employed the relation
\begin{equation}
\label{Sdiff}
 -\frac{1}{2k}\bigl( S(k)-S^\ast (k)\bigr) = -2\ui \frac{L}{L^2 +k^2} 
\end{equation}
that follows from \eqref{SthruPQ}. This then yields
\begin{equation}
\label{kderiv1}
\begin{split}
\left<v(k),\dk[U(k)v(k)]\right> 
      &= - 2\ui\,\ue^{\ui \theta(k)}\left<v(k),\frac{L}{L^2+k^2}v(k)\right> 
             +\ui \left<U^{\ast}(k)v(k),D(\vl)v(k)\right> \\
      &\quad +\left<U^{\ast}(k)v(k),v'(k)\right>\\ \\
      &= -2\ui\,\ue^{\ui \theta(k)}\left<v(k),\frac{L}{L^2+k^2}v(k)\right> 
            + \ui\,\ue^{\ui \theta(k)}\left<v(k),D(\vl)v(k)\right> \\
      &\quad + \ue^{\ui\theta(k)}\left<v(k),v'(k)\right> \ .
\end{split}
\end{equation}
On the other hand, taking the derivative on the right-hand side of 
(\ref{15}) and then multiplying with $v(k)$ leads to
\begin{equation}
\label{kderiv2}
 \left<v(k),\dk [\ue^{\ui\theta(k)}v(k)]\right> =\ui\theta'(k)\ue^{\ui\theta(k)}
 +\ue^{\ui \theta(k)}\left<v(k),v'(k)\right> \ .
\end{equation}
Comparing \eqref{kderiv1} and \eqref{kderiv2} proves the statement 
\eqref{kderivlem}.                             
\end{proof}
This lemma allows us to obtain an upper and a lower bound for the derivative
of an eigenphase. Using $l_{\max/\min}$ to denote the largest and the smallest 
edge length, respectively, and  introducing
\begin{equation*}
 \lmm := \begin{cases}
	     \min\{|\lambda|;\ \lambda\in\sigma(L)\cap\rz^-\}, & \quad
             \text{if}\quad \exists\lambda_\alpha < 0 \\
	     \infty, & \quad \text{else}
	 \end{cases} \ ,
\end{equation*}
in analogy to \eqref{lambdaminplus}, we immediately get the following.
\begin{cor}
\label{21}
The derivative $\theta'(k)$ of an eigenphase $\theta(k)$ is bounded from
above and below according to
\begin{equation}
\label{phasederest}
 l_{\min}-\frac{2}{\lmp} \leq \theta'(k) \leq l_{\max}+\frac{2}{\lmm} \ .
\end{equation}
In particular, if $l_{\min}>2/\lmp$ the derivatives of all eigenphases
are always positive.
\end{cor}
\begin{proof}
Obviously,
\begin{equation*}
 l_{\min}\leq\sum\limits_{i=1}^{2E}l_i|v_i(k)|^2\leq l_{\max} \ ,
\end{equation*}
since the eigenvector is supposed to be normalised. Moreover,
after a diagonalisation of $L$, when $WLW^\ast$ is diagonal with
the eigenvalues $\lambda_\alpha$ on the diagonal, one obtains
\begin{equation*}
 \left<v(k),\frac{L}{L^2+k^2}v(k)\right> =  \sum_{\alpha=1}^{2E}
 |w_\alpha(k)|^2\frac{\lambda_\alpha}{\lambda^2_\alpha+k^2} \ , 
\end{equation*}
where $w(k)=Wv(k)$. This first yields
\begin{equation*}
 -\frac{1}{\lmm}\leq\left<v(k),\frac{L}{L^2+k^2}v(k)\right>\leq
 \frac{1}{\lmp} \ ,
\end{equation*}
and then finally \eqref{phasederest}.
\end{proof}
As an important consequence of this corollary we are now able to extend
the statement of Proposition~\ref{thm:seceq} as required to count eigenvalues
in terms of zeros of the determinant function. 
\begin{prop}
\label{27}
Let the metric structure of the graph be such that $l_{\min}>2/{\lmp}$. Then
$k_n^2>0$ is an eigenvalue of the Laplacian with multiplicity $g_n$, if
$\pm k_n\in\rz\setminus\{0\}$ are zeros of the function $F(k)$ of order
$g_n$.
\end{prop}
\begin{proof}
From Proposition~\ref{thm:seceq} and Lemma~\ref{functeq} we know that the 
positive Laplace eigenvalues are in one-to-one correspondence with (pairs of)
$k$ values where $U(k)$ has an eigenvalue one, and the respective 
multiplicities coincide. It hence remains to establish that the order of
the corresponding zeros of $F(k)$ are exactly these multiplicities: 
From the definition \eqref{23} of the function $F(k)$ one obtains
\begin{equation*}
 F(k)=\prod_{j=1}^{2E}\left(1-\ue^{\ui \theta_j(k)}\right) \ ,
\end{equation*}
so Corollary~\ref{21} implies $\dk\left(1-\ue^{\ui \theta_j(k)}\right)=
-\ui\theta_j'(k)\ue^{\ui\theta_j(k)}\neq0$, and the claim follows immediately.
\end{proof}
We remark that since the scattering approach to the proof of the trace
formula is based on counting zeros of the function $F(k)$ on the real
line with their multiplicities, the requirement $l_{\min}>2/{\lmp}$ is 
essential. Otherwise one might count Laplace eigenvalues with incorrect
multiplicities. Whenever $L$ has no positive part, however, the condition
is empty. This is, e.g., the case for non-Robin boundary conditions.
\subsection{The eigenvalue zero}
In general, zero is a Laplace eigenvalue as well as a zero of the 
determinant~\eqref{23}, and in so far Proposition~\ref{thm:seceq} also applies 
to $k_0 =0$. The spectral multiplicity $g_0$, however, typically is different 
from the degree of $k_0 =0$ as a zero of $F(k)$. For Kirchhoff boundary 
conditions it has been shown in \cite{Kurasov:2005} that the degree of the zero
is $E-V+2$, whereas the zero Laplace eigenvalue is non-degenerate, i.e., 
$g_0 =1$. Kurasov \cite{Kurasov:2007} subsequently linked this difference in 
the multiplicities to the topology of the graph by noticing that a suitable 
trace formula contains the quantity
\begin{equation*}
 1 -  \frac{1}{2}(E-V+2) = \frac{1}{2}(V-E) \ ,
\end{equation*} 
and hence the Euler characteristic of the graph. This observation was
generalised to yield an index theorem for any quantum graph with non-Robin
boundary conditions by Fulling, Kuchment and Wilson \cite{Fulling:2007}. One 
can view the $t$-independent term $\tfrac{1}{4}\mtr S = \tfrac{1}{2}(V-E) $ 
in the trace formula for the heat kernel (due to \cite{Roth:1983} for 
Kirchhoff boundary conditions and \cite{Schrader:2007} for general non-Robin 
conditions) as a predecessor of this result. See also \cite{BolteEndres:2007} 
for a more detailed discussion.
  
We here wish to give a further characterisation of the spectral multiplicity
of the zero eigenvalue in the case of a general self adjoint realisation of
the Laplacian. To this end we first introduce, for $k\in\rz\setminus\{0\}$,
the matrix
\begin{equation}
\label{def:Cmatrix}
 C(\vl;k) := \begin{pmatrix}\begin{matrix}
                                              \frac{l_1}{2\frac{\ui}{k} +l_1} & & \\
                                               & \ddots & \\
                                               & & \frac{l_E}{2\frac{\ui}{k} +l_E}
                                              \end{matrix} &
                                              \begin{matrix}
                                              \frac{2\frac{\ui}{k}}{2\frac{\ui}{k} +l_1} & & \\
                                               & \ddots & \\
                                               & & \frac{2\frac{\ui}{k}}{2\frac{\ui}{k} +l_E}
                                              \end{matrix} \\
                                              \begin{matrix}
                                              \frac{2\frac{\ui}{k}}{2\frac{\ui}{k} +l_1} & &  \\
                                               & \ddots & \\
                                               & & \frac{2\frac{\ui}{k}}{2\frac{\ui}{k} +l_E}
                                              \end{matrix} &
                                              \begin{matrix}
                                              \frac{l_1}{2\frac{\ui}{k} +l_1} & & \\
                                               & \ddots & \\
                                               & & \frac{l_E}{2\frac{\ui}{k} +l_E}
                                              \end{matrix}
                      \end{pmatrix} \ ,
\end{equation}    
in which all matrix entries not indicated are zero. This now enables us to 
formulate the following.
\begin{prop}
\label{64}
For any given self adjoint realisation of the Laplacian specified through $A,B$,
zero is a Laplace eigenvalue, iff one is an eigenvalue of $S(A,B;k)C(\vl;k)$ 
for one, and hence any, $k\in\rz\setminus\{0\}$. Moreover, the multiplicity of 
this eigenvalue one coincides with the spectral multiplicity $g_0$ of the zero
Laplace eigenvalue.
\end{prop}
\begin{proof}
Eigenfunctions of the Laplacian corresponding to the eigenvalue zero must be
of the form $F=(f_1,\dots,f_E)^T$ with
\begin{equation}
\label{0eigenfct}
 f_j(x) = \alpha_j+\beta_j x \ ,\quad x\in[0,l_j] \  .
\end{equation}
Hence, the boundary values (\ref{def:bv}) take the form
\begin{equation*}
\begin{split}
 F_{bv}  &= \bigl(\alpha_1,\dots,\alpha_E,\alpha_1+\beta_1l_1,\dots,\alpha_E
                    +\beta_1l_E \bigr)^T \ , \\
 F'_{bv}  &= \bigl( \beta_1,\dots,\beta_E,-\beta_1,\dots,-\beta_E \bigr)^T \ .
\end{split}
\end{equation*} 
We now employ the boundary conditions (\ref{bc}) by using the expressions 
\eqref{ABprime} for a possible choice of $A$ and $B$. The result can be
rearranged to yield
\begin{equation}
\label{206}
 S(k)C_+(\vl;k) \begin{pmatrix}\valpha \\ \vbeta \end{pmatrix}
      = C_-(\vl;k)\begin{pmatrix} \valpha \\ \vbeta \end{pmatrix} \ ,
\end{equation}
where, for any $k\in\rz\setminus\{0\}$, we have introduced
\begin{equation*}
 C_{\pm}(\vl;k) := \begin{pmatrix} 
       \eins_E & \pm\frac{\ui}{k} \eins_E \\
       \eins_E & \mp\frac{\ui}{k} \eins_E +D_1(\vl) \end{pmatrix} \ ,
\end{equation*}
with $D_1(\vl)$ as defined in \eqref{diagL}. We also use the abbreviations
\begin{equation*}
 \valpha := \begin{pmatrix} \alpha_1 \\ \vdots \\ \alpha_E
 \end{pmatrix} \quad \text{and} \quad
 \vbeta := \begin{pmatrix} \beta_1 \\ \vdots \\ \beta_E
 \end{pmatrix} \ .
\end{equation*}
The matrices $C_{\pm}(\vl,k)$ are invertible for all $k\in\rz\setminus\{0\}$, with
\begin{equation*}
 C_\pm(\vl;k)^{-1} = \begin{pmatrix}
     \begin{matrix} \frac{\pm \frac{\ui}{k} -l_1}{\pm2\frac{\ui}{k}-l_1} & & \\
                               & \ddots & \\
                               & & \frac{\pm \frac{\ui}{k} -l_E}{\pm2\frac{\ui}{k}-l_E}
     \end{matrix} &
     \begin{matrix} \frac{\pm \frac{\ui}{k}}{\pm2\frac{\ui}{k}-l_1} & & \\
                               & \ddots & \\
                               & & \frac{\pm \frac{\ui}{k}}{\pm2\frac{\ui}{k}-l_1}
     \end{matrix} \\
     \begin{matrix} \frac{1}{\pm2\frac{\ui}{k}-l_1} & & \\
                               & \ddots & \\
                               & & \frac{1}{\pm2\frac{\ui}{k}-l_E}
     \end{matrix} &
     \begin{matrix} \frac{-1}{\pm2\frac{\ui}{k}-l_1} & & \\
                               & \ddots & \\
                               & & \frac{-1}{\pm2\frac{\ui}{k}-l_1}
     \end{matrix}  \end{pmatrix} \ .
\end{equation*}
We now substitute 
\begin{equation}
\label{210}
 v(k):=C_{-}(\vl;k)\begin{pmatrix}\valpha \\ \vbeta \end{pmatrix}
\end{equation}
in \eqref{206} and obtain 
\begin{equation*}
 S(k)C_+(\vl;k)C_-(\vl;k)^{-1}v(k) = v(k) \ .
\end{equation*}
It is straight forward to check that $C_+(\vl;k)C_-(\vl;k)^{-1}=C(\vl;k)$, 
compare \eqref{def:Cmatrix}. The linearly independent eigenvectors of $SC$ 
corresponding to the eigenvalue one then yield, via \eqref{210} and 
\eqref{0eigenfct}, coefficients $\alpha_j$ and $\beta_j$ for as many linearly 
independent Laplace eigenfunctions in $L^2(\Gamma)$.
\end{proof}
We remark that the order $N$ of $k_0 =0$ as a zero of the function \eqref{23} 
is the multiplicity of the eigenvalue one of 
\begin{equation*}
 U(0) = S_0 \, \begin{pmatrix} 0 & \eins_E \\ \eins_E & 0 \end{pmatrix} \ ,
\end{equation*}
which is in no obvious way related to the multiplicity of the eigenvalue one of
$S(k)C(\vl;k)$ that appears in Proposition \ref{64}. In the case of non-Robin
boundary conditions, where the edge S-matrix is independent of $k$, however,
Fulling, Kuchment and Wilson \cite{Fulling:2007} were able to relate the 
different multiplicities in the form of an index theorem. They showed, in 
particular, that then
\begin{equation*}
 g_0 - \frac{1}{2}N = \frac{1}{4}\mtr S \ .
\end{equation*}
As mentioned above, this term will reappear in the trace formula.
\section{The trace formula}
\label{prw}
A trace formula expresses counting functions of Laplace eigenvalues
in terms of sums over periodic orbits. Ideally, one would like to count
Laplace eigenvalues $k_n^2$, with their multiplicities $g_n$, in
intervals $I$ in the form
\begin{equation}
\label{sharpcount}
 \tr \chi_I (-\Delta) = \sum_{k_n^2\in I}g_n \ .
\end{equation}
The sharp cut-off provided by the characteristic function $\chi_I$ of
the interval $I$, however, cannot be dealt with. One therefore replaces
\eqref{sharpcount} with a smooth cut-off and, moreover, performs this
count in terms of the associated wave numbers $k_n$, i.e., one seeks
to find a representation for 
\begin{equation}
\label{lhsTF}
 \sum_n g_n \, h(k_n)
\end{equation}
in terms of sums over periodic orbits. One ambition then is to find a 
sufficiently large class of test functions $h$. It turns out that the
following one parameter family of test functions is particularly suited
for these purposes.
\begin{defn}
\label{def:Hr}
For each $r\geq 0$ the space $H_r$ consists of all functions
$h:\kz\to\kz$ satisfying the following conditions:
\begin{itemize}
\item $h$ is even, i.e., $h(k)=h(-k)$. 
\item For each $h\in H_r$ there exists $\delta>0$ such that $h$ is analytic 
in the strip $M_{r+\delta}:=\{k\in\kz; \ |\im k|<r+\delta\}$.
\item For each $h\in H_r$ there exists $\eta>0$ such that 
$h(k)=O\left(\frac{1}{(1+|k|)^{1+\eta}}\right)$ on $M_{r+\delta}$.
\end{itemize}
\end{defn}
We stress that the trace formula will only take non-negative Laplace 
eigenvalues into account. Hence, the test functions in \eqref{lhsTF} will be 
evaluated at real arguments. We can therefore arrange the wave numbers $k_n>0$ 
corresponding to Laplace eigenvalues $k_n^2$ in ascending order.
Proposition~\ref{Weylnew} then readily implies the existence of a constant 
$C>0$ such that
\begin{equation}
\label{specsum}
 \sum_{0<k_n\leq K} g_n \, |h(k_n)| \leq C K \,\underset{k\in[0,K]}{\sup}
 |h(k)| \ .
\end{equation}
The third condition of Definition~\ref{def:Hr} now ensures that the sum in
\eqref{lhsTF} converges absolutely when $h\in H_r$ with some $r\geq 0$.
Since this sum constitutes one side of the trace formula, choosing test
functions $h$ from any of the spaces $H_r$ is an appropriate prerequisite
for the trace formula.
\subsection{A precise eigenvalue count and the trace formula}
The trace formula is based on an exact count of non-negative Laplace 
eigenvalues with, however, a smooth cut-off as in \eqref{lhsTF}. Essentially, 
this sum is expressed in terms of a sum over periodic orbits on the graph. 
In order to perform such an eigenvalue count one relies on the connection 
between Laplace eigenvalues and zeros of the determinant function \eqref{23}. 
More specifically,  one chooses a test function $h\in H_r$ with some $r\geq 0$ 
as well as some $\varepsilon>0$, which must be sufficiently small so that in 
the set 
$C_{\varepsilon,K}=\{ k\in\kz ;\ |\im k|\leq\varepsilon,\ |\re k| \leq K \}$
the determinant function $F(k)$ has only (finitely many) real zeros related 
to non-negative Laplace eigenvalues (compare Proposition~\ref{thm:seceq}). 
Then, by the argument principle,
\begin{equation}
\label{finK}
 \frac{1}{2\pi\ui}\int_{\partial C_{\varepsilon,K}} \frac{F'}{F}(k)\,h(k)\ \ud k =
 N\,h(0) + 2\sum_{0<k_n\leq K} g_n \, h(k_n) \ .
\end{equation}
Here $N$ and $g_n$ are the multiplicities of $k_0 =0$ and $k_n >0$, 
respectively, as a zeros of $F(k)$. Following Proposition~\ref{27},
$g_n$ also is the multiplicity of the Laplace eigenvalue $k_n^2$. The
factor of two occurs since $k_n,-k_n\in C_{\varepsilon,K}$ and we have 
exploited the fact that $h(k)$ is even. Based on this relation the trace 
formula emerges when one expresses $F'(k)/F(k)$ in terms of a suitable 
series (eventually leading to a sum over periodic orbits), and performs the 
limit $K\to\infty$. The result of this procedure is summarised in the 
following statement.
\begin{prop}
Let $l_{\min}>2/\lmp$ and choose $h\in H_r$ with any $r\geq0$. Then
\begin{equation}
\label{pretrace}
 N\,h(0) + 2\sum_{n=1}^\infty g_n \, h(k_n)  
  =\sum_{l\in\gz} \frac{1}{2\pi\ui}\int_{-\infty}^{+\infty}\mtr \bigl[ 
  \Lambda(k)U^l (k) \bigr] \, h(k) \ \ud k \ ,
\end{equation}
where 
\begin{equation}
\label{Lambdadef}
 \Lambda(k)=-\ui\frac{2L}{L^2+k^2}+\ui D(\vl) \ .
\end{equation}
\end{prop}
\begin{proof}
Our strategy is to show that both sides of \eqref{pretrace} equal
\begin{equation}
\label{zerocountint}
 \lim_{\varepsilon\to 0^+}
 \frac{1}{2\pi\ui}\int_{-\infty}^{+\infty} \left[ \frac{F'}{F}(k-\ui\varepsilon)
 \,h(k-\ui\varepsilon) -\frac{F'}{F}(k+\ui\varepsilon)\,h(k+\ui\varepsilon) 
  \right]\,\ud k \ .
\end{equation}
Beginning with the left-hand side, we have to show that in \eqref{finK} the 
limit $K\to\infty$ can be taken, followed by $\varepsilon\to 0$. To this end 
we notice that the right-hand side of \eqref{finK} is explicitly independent 
of $\varepsilon$, in the range described above that equation. Furthermore, 
from \eqref{specsum} we already know that the sum over $k_n$ converges 
absolutely in the limit $K\to\infty$. In order to perform these limits on
the left-hand side of \eqref{finK}, and thus producing \eqref{zerocountint},
we have to estimate the contribution to the integral coming from the
vertical parts of the contour.

Lemma~\ref{70} implies that for $k\in\kz$ with $|k|>\lambda_{\max}$ the 
approximation $F(k) = F_\infty (k) + O(|k|^{-1})$ holds, where
\begin{equation}
\label{Finftydef}
 F_\infty (k) := \det\bigl(\eins - S_\infty T(k) \bigr) \ .
\end{equation}
Since $F_\infty$ depends on $k$ only through the matrix entries $\ue^{\ui kl_e}$
of $T(k)$, see \eqref{def:Tofk}, it can be represented as 
\begin{equation}
\label{Finfty}
 F_\infty (k) = 1 + \sum_{n=1}^b d_n \, \ue^{\ui\beta_n k} \ ,
\end{equation}
with some $b<\infty$. Moreover, $\beta_n>0$ is a finite sum of edge lengths, 
and $d_n \in\kz$ is an appropriate coefficient. The expression \eqref{Finfty}
is first defined for $k\in\rz$, but can be readily extended to complex $k$.
Since $S'(k)= O(|k|^{-2})$, see Lemma~\ref{16} and eq.~\eqref{Sdiff}, we also 
find $F'(k)=F_\infty' (k)+O(|k|^{-1})$. For sufficiently large $|k|$ one can 
therefore approximate $F'(k)/F(k)$ by $F_\infty' (k)/F_\infty (k)$. In order 
to estimate the latter we employ \eqref{Finfty} to obtain
\begin{equation}
\label{Fderest}
 \bigl| F_\infty' (k) \bigr| \leq  b \, d_{\max} \, \beta_{\max} \, 
 \ue^{-\beta_{\min} \im k} \ ,
\end{equation}
with $d_{\max}:=\max\{|d_n|\}$ and $\beta_{\max/\min}:=\max/\min\{\beta_n\}$.
Note that this bound is independent of $\re k$. Furthermore, we pick 
$k_{(0)}\in\rz$ such that $F(k_{(0)})\neq 0$ as well as 
$F_\infty(k_{(0)})\neq 0$, and take advantage of the fact that 
$F_\infty(k)$, $k\in\rz$, is an almost periodic function (see, e.g., 
\cite{bohr:1947}). One can hence construct
a (strictly increasing) sequence $\{k_{(j)} ;\ j\in\nz_0\}$ with
\begin{equation}
 |F_\infty (k_{(j)}) - F_\infty (k_{(0)})| < \frac{|F_\infty (k_{(0)})|}{2} \ ,
 \quad j>0 \ .
\end{equation}
Hence, in particular, $F_\infty(k_{(j)})\neq 0$. The estimate \eqref{Fderest}, 
moreover, implies that 
\begin{equation}
 |F_\infty (k_{(j)}) - F_\infty (k_{(j)} +\ui\kappa)| < 
 \frac{|F_\infty (k_{(0)})|}{4} 
\end{equation}
when $|\kappa|$ is sufficiently small. Therefore, when $\varepsilon$ is small 
enough, the function $|F_\infty (k_{(j)} +\ui\kappa)|$, 
$-\varepsilon\leq\kappa\leq\varepsilon$, is uniformly bounded from below away 
from zero. Thus there exists $C_\varepsilon >0$ such that
\begin{equation}
 \left| \int_{k_{(j)}-\ui\varepsilon}^{k_{(j)} +\ui\varepsilon}
 \frac{F_\infty' (k)}{F_\infty (k)}\,h(k) \ \ud k \right| \leq C_\varepsilon
 \,\sup \{|h(k_{(j)} +\ui\kappa)|;\ \kappa\in (-\varepsilon,+\varepsilon)\} \ .
\end{equation}
The third property of test functions required in Definition~\ref{def:Hr} 
then ensures that this integral vanishes as $k_{(j)}\to\infty$. A completely
analogous reasoning applies when replacing $k_{(j)}$ with $-k_{(j)}$. Hence, as
$K\to\infty$ the contributions coming from 
$[-K+\ui\varepsilon,-K-\ui\varepsilon]$ and 
$[K-\ui\varepsilon,K+\ui\varepsilon]$ to the integration along 
$\partial C_{\varepsilon,K}$ in \eqref{finK} vanish, and the limits
$\varepsilon\to 0$ and $K\to\infty$ can be interchanged.

It remains to prove the equality of \eqref{zerocountint} with the 
right-hand side of \eqref{pretrace}. In order to achieve this we first recall 
from Proposition~\ref{thm:seceq} that the function $F(k)$ is holomorphic
and zero-free in the strip $\{ k\in\kz;\ 0< \im k < \varepsilon_1\}$, 
where $\varepsilon_1 := \min\{ \lmp,\kappa_1 \}$ when $-\kappa_1^2$ 
is the largest negative eigenvalue of $-\Delta$; if $-\Delta\geq 0$ we
set $\varepsilon_1 := \lmp$. Hence,
\begin{equation*}
 \frac{F'}{F}(k+\ui\varepsilon)  = \dk\log\det\bigl(\eins - 
 U(k+\ui\varepsilon)\bigr) = 
 -\mtr\Bigl(\bigl(\eins - U(k+\ui\varepsilon)\bigr)^{-1}
 U'(k+\ui\varepsilon)\Bigr) \ ,
\end{equation*}
when $0<\varepsilon<\varepsilon_1$. Furthermore, following 
Lemma \ref{Sbound2} we obtain the bound
\begin{equation}
\label{Unorm}
 \| U(k+\ui\varepsilon) \| \leq 
 \| S(k+\ui\varepsilon) \| \| T(k+\ui\varepsilon) \| \leq
 \max\left\{ 1,\frac{\lmp+\varepsilon}{\lmp -\varepsilon} \right\} \, 
 \, \ue^{-\varepsilon l_{\min}} 
\end{equation}
in the operator norm. Thus, when
\begin{equation}
\label{epscond}
 \varepsilon < \lmp \tanh\Bigl(\frac{\varepsilon l_{\min}}{2}\Bigr) \ ,
\end{equation}
the operator norm of $U(k+\ui\varepsilon)$ is less than one. Iff 
$l_{\min}>2/\lmp$ this condition is solvable, providing some $\varepsilon_2$ 
with $0<\varepsilon_2<\lmp$ such that \eqref{epscond} is true for all 
$\varepsilon\in (0,\varepsilon_2)$. For such $\varepsilon$ the expansion
\begin{equation*}
 \bigl(\eins - U(k+\ui\varepsilon)\bigr)^{-1} =
 \sum_{l=0}^\infty U(k+\ui\varepsilon)^l
\end{equation*}
holds. Moreover, a straight forward calculation based on the relation
\eqref{Udash} produces $\mtr [U(k)^lU'(k)]=\mtr[\Lambda(k)U(k)^{l+1}]$, see 
\eqref{Lambdadef}. Therefore, after cyclic permutations under the trace,
\begin{equation}
\label{logFder1}
 \frac{F'}{F}(k+\ui\varepsilon)  = -\sum_{l=1}^\infty \mtr\bigl(  
 \Lambda(k+\ui\varepsilon)  U^l(k+\ui\varepsilon)  \bigr) 
\end{equation}
is true for all $k\in\rz$ and $\varepsilon\in (0,\varepsilon_3)$,
where $\varepsilon_3 := \min\{\varepsilon_1,\varepsilon_2\}$. 

Likewise, based on the relation
\begin{equation*}
 \bigl(\eins - U(k-\ui\varepsilon)\bigr)^{-1} = 
 -\bigl(\eins - U^{-1}(k-\ui\varepsilon)\bigr)^{-1} 
 U^{-1}(k-\ui\varepsilon) 
\end{equation*}
that holds for $0<\varepsilon<\lmp$, we obtain
\begin{equation}
\label{logFder2}
 \frac{F'}{F}(k-\ui\varepsilon)  = \sum_{l=0}^\infty \mtr\bigl(  
 \Lambda(k-\ui\varepsilon)  U^{-l}(k-\ui\varepsilon)  \bigr) 
\end{equation}
for $k\in\rz$ and sufficiently small $\varepsilon>0$.

We now want to use the representations \eqref{logFder1} and \eqref{logFder2} 
in \eqref{zerocountint}, interchange integration and summations, and
finally perform the limit $\varepsilon\to 0$. In order to achieve this we
first notice that (for fixed $\varepsilon$) the estimate
\begin{equation}
\label{normesti}
\begin{split}
 \bigl| \mtr\bigl(  \Lambda(k\pm\ui\varepsilon) U^{\pm l}(k\pm\ui\varepsilon)  
 \bigr)  \bigr| 
 &\leq 2E \|\Lambda(k\pm\ui\varepsilon) U^{\pm l}(k\pm\ui\varepsilon) \| \\
 &\leq 2E \| \Lambda(k\pm\ui\varepsilon) \| 
  \| U^{\pm 1}(k\pm\ui\varepsilon) \|^l 
\end{split}
\end{equation}
justifies to interchange integration and summation according to the dominated
convergence theorem. Next, the contour of the integral
\begin{equation}
\label{epsint}
 \int_{\im k =\pm\varepsilon} \mtr\bigl[ \Lambda(k)U^{\pm l}(k) \bigr] \, 
 h(k)\  \ud k  
\end{equation}
can be deformed into $\im k =\pm\delta$ with a sufficiently small $\delta>0$; 
in particular, $\delta<r$. We then fix $\delta$ and find
\begin{equation}
\label{lestimate}
\begin{split}
 \Bigl|\int_{\im k =\pm\varepsilon} 
 &\mtr\bigl[ \Lambda(k)U^{\pm l}(k) \bigr] \, h(k)\  \ud k \Bigr| \\
 &\leq \int_{-\infty}^{+\infty} \bigl| \mtr\bigl(  \Lambda(k\pm\ui\delta)  
  U^{\pm l}(k\pm\ui\delta)  \bigr)  \bigr| \, | h(k\pm\ui\delta) | \ud k \\
 &\leq 2E \left(  \frac{\lmp+\delta}{\lmp -\delta} \, 
  \ue^{-\delta l_{\min}} \right)^l \int_{-\infty}^{+\infty} 
  \| \Lambda(k\pm\ui\delta) \| \, | h(k\pm\ui\delta) | \ud k \ ,
\end{split}
\end{equation}
when using \eqref{Unorm} and \eqref{normesti} with $\delta$ instead of 
$\varepsilon$. Since the conditions in Definition~\ref{def:Hr} apply, the 
integral on the right-hand side is finite and the positive constant raised 
to the power $l$ is smaller than one. The sum on $l$ hence possesses an 
absolutely convergent majorant, uniform in $\varepsilon$ in the range 
indicated. Thus the summation and the limit $\varepsilon\to 0$ can be 
interchanged. Furthermore, another application of the dominated convergence 
theorem allows to perform the limit $\varepsilon\to 0$ of \eqref{epsint}  
inside the integral, finally proving \eqref{pretrace}.
\end{proof}
In order to arrive at the trace formula itself, the sum  on the right-hand
side of \eqref{pretrace} has to be reformulated as a sum over periodic 
orbits. The summation index $l$ then denotes the topological length of
the orbits and the trace of $\Lambda(k) U^l (k)$ is identified as a sum 
over the set $\cP_l$ of periodic orbits with topological length $l$. The 
term with $l=0$ plays a special role and will be treated separately,
whereas contributions with negative $l$ are related to those with $-l$
in a simple way. 

Before we state the trace formula, however, we introduce some notation:
The `volume' of the graph is the sum $\cL = l_{e_1}+\dots +l_{e_E}$ of 
all edge lengths. Furthermore, if $h\in H_r$ is a test function, its Fourier 
transform is 
\begin{equation*}
 \hat h (x) = \frac{1}{2\pi}\int_{-\infty}^{+\infty} h(k) \, \ue^{\ui kx} 
 \  \ud k \ .
\end{equation*}
We recall that the second and the third property required for $h$ in 
Definition~\ref{def:Hr} guarantee that $\hat h(x) = O(\ue^{-rx})$ as 
$x\to\infty$. Moreover, the Fourier transform of a product $A(k)h(k)$ is the 
convolution of the respective Fourier transforms, i.e., it reads
\begin{equation}
\label{convol}
 \widehat{Ah} (x) = \int_{-\infty}^{+\infty}\hat A (x-y)\, \hat h(y) \  \ud y
 = \hat A * \hat h (x) \ .
\end{equation}
Below this convolution will often have to be understood in a distributional 
sense, as the functions $A(k)$, though being regular, not always decay 
sufficiently fast as $k\to\infty$.

We are now in a position to state the first variant of the trace formula.
\begin{theorem}
\label{74}
Let $\Gamma$ be a compact, metric graph with a self adjoint realisation 
$-\Delta(A,B;\vl)$ of the Laplacian, such that the condition 
$l_{\min}>2/\lmp$ is fulfilled. Furthermore, let $h\in H_r$ be a test function
with an arbitrary $r\geq 0$. Then the following identity holds:
\begin{equation}
\begin{split}
\label{TF1}
 \sum_{n=0}^{\infty} g _n \, h(k_n) 
 &= \phantom{+}\cL \, \hat{h}(0) + \bigl(g_0 -\tfrac{1}{2}N \bigr) \, h(0)
    -\frac{1}{4\pi}\int_{-\infty}^{+\infty}h(k)\,\frac{\im\mtr S(k)}{k}
    \ \ud k \\
 &  \quad +\sum_{l=1}^{\infty}\sum_{p\in\cP_l}\left[\bigl(\hat{h}\ast\hat{A}_p
    \bigr) (l_p) + \bigl(\hat{h}\ast\hat{\overline{A}}_p\bigr)(l_p) \right]
     \ .
\end{split}
\end{equation}
Here $\hat{A}_p$ is the Fourier transform of the amplitude function $A_p(k)$
associated with every periodic orbit $p$. This function is meromorphic with
possible poles at the poles of the S-matrix, and has a Taylor expansion
\begin{equation}
\label{amplex}
 A_p (k) = \sum_{j=0}^\infty a_p^{(j)} k^{-j} 
\end{equation}
that converges for $|k|>\lambda_{\max}$. In general, only the sum over the
topological lengths $l$ on the right-hand side of \eqref{TF1} converges
absolutely, but not the entire double sum over periodic orbits.
\end{theorem}
\begin{proof}
We have to evaluate $\mtr [\Lambda(k)U^l (k)]$, and first notice that
according to \eqref{def:Tofk} and \eqref{def:Uofk} the matrix entries of 
$U$ are
\begin{equation*}
 U_{j_1 j_2} = S_{j_1\omega_{j_2}}\, \ue^{\ui kl_{j_1}} \ ,                          
\end{equation*} 
where
\begin{equation*}
 \omega_j := \begin{cases} j+E, & \text{if}\ 1\leq j\leq E \\
                   j-E, & \text{if}\ E+1 \leq j\leq 2E  \end{cases} \ .     
\end{equation*} 
Hence, the indices $j$ and $\omega_j$ correspond to the two edge ends of
the $j$-th edge. Local boundary conditions then imply that 
$S_{j_1\omega_{j_2}}\neq 0$ requires the edges with ends $j_1$ and $\omega_{j_2}$
to be adjacent. Therefore, when $l>0$ the non vanishing terms in the multiple 
sum
\begin{equation}
\label{ltrace}
 \mtr\bigl( \Lambda U^l \bigr) = \sum_{j_0,\dots,j_l =1}^{2E} \Lambda_{j_0 j_1}
 S_{j_1\omega_{j_2}}\dots S_{j_l\omega_{j_0}} \, 
 \ue^{\ui k (l_{j_1} +\cdots + l_{j_l})} 
\end{equation} 
correspond to the closed paths of topological length $l$ on the graph.

We then make use of the decomposition \eqref{Lambdadef} of $\Lambda$ and begin
with the contribution of $\mtr [D(\vl)U^l(k)]$, which can be evaluated as in 
the case of Kirchhoff boundary conditions \cite{KottosSmilansky:1998}:
Due to the specific diagonal form of $D$ the terms in \eqref{ltrace} 
corresponding to closed paths related by cyclic permutations of their edges 
can be grouped together. This finally yields a sum over the periodic orbits
of topological length $l$,
\begin{equation*}
 \mtr\bigl[D(\vl)U^l(k)\bigr] = 2\sum_{p\in\cP_l}l_p^\# A_{1,p}(k)\,
 \ue^{\ui k l_p} \ .                            
\end{equation*} 
According to \eqref{ltrace}, the functions $A_{1,p}(k)$ result from multiplying
the local S-matrices of the vertices visited along the periodic orbit $p$.
Moreover, $l_p^\#$ is the primitive length of $p$, i.e., the length
of the primitive periodic orbit associated with $p$. Due to the relation 
\eqref{convol} we therefore obtain
\begin{equation*}
 \frac{1}{2\pi}\int_{-\infty}^{+\infty} \mtr [D(\vl)U^l(k)]\,h(k) \ \ud k
 =  2\sum_{p\in\cP_l}l_p^\# \bigl(\hat{h}\ast\hat{A}_{1,p} \bigr)(l_p) \ .
\end{equation*}
The case of negative $l$ follows by noticing that 
\begin{equation*}
 U^{-1}(k) = T^{-1}(k)S^{-1}(k) = T(-k)S(-k) \ , 
\end{equation*}
and thus 
\begin{equation*}
 \mtr\bigl[D(\vl)U^{-l}(k)\bigr] = 2\sum_{p\in\cP_l}l_p^\#\overline{A}_{1,p}(k)\,
 \ue^{-\ui k l_p} \ ,  
\end{equation*}
leading to
\begin{equation*}
\begin{split}
 \frac{1}{2\pi}\int_{-\infty}^{+\infty} \mtr\left[D(\vl)\bigl(U^l(k)\right.
 &+ \left.U^{-l}(k)\bigr) \right]\,h(k) \ \ud k \\
 &=  2\sum_{p\in\cP_l}l_p^\# \left[ \bigl(\hat{h}\ast\hat{A}_{1,p} \bigr)(l_p) +
    \bigl(\hat{h}\ast\hat{\overline{A}}_{1,p} \bigr)(l_p) \right] \ .
\end{split}
\end{equation*}
In order to calculate the contribution from $\mtr [\frac{L}{L^2+k^2}U^l(k)]$
we notice that
\begin{equation}
\label{ltrace2}
 -2\ui \mtr \left[\frac{L}{L^2+k^2}U^l \right] = \mtr [ S'T U^{l-1} ] = 
 \sum_{j_1,\dots,j_l =1}^{2E} S'_{j_1\omega_{j_2}} S_{j_2\omega_{j_3}}\dots 
 S_{j_l\omega_{j_1}} \, \ue^{\ui k (l_{j_1} +\cdots + l_{j_l})} \ .
\end{equation} 
Again, the multiple sum can be viewed as a sum over the closed paths of
topological length $l$, and the contributions of representatives of periodic
orbits can be grouped together. Eventually, this leads to
\begin{equation}
\label{Lcontrib}
 -2\mtr\left[ \frac{L}{L^2+k^2}U^l(k) \right] = 
 2\sum_{p\in\cP_l}A_{2,p}(k)\,\ue^{\ui k l_p} \ ,                           
\end{equation} 
where the functions $A_{2,p}(k)$ emerge from multiplying local S-matrix
elements and their derivatives along the closed paths as specified in
\eqref{ltrace2}. Negative $l$ are dealt with as above, so that the 
contribution from $l\in\gz\setminus\{0\}$ to the sum on the right-hand side 
of \eqref{pretrace} yields the sum on the right-hand side of \eqref{TF1},
with $A_p (k)=l_p^\# A_{1,p}(k) + A_{2,p}(k)$.

The contribution coming from $l=0$ finally is
\begin{equation*}
\begin{split}
 \frac{1}{2\pi\ui}\int_{-\infty}^{+\infty} \mtr\Lambda(k)\,h(k) \ \ud k
  &= -\frac{1}{2\pi}\int_{-\infty}^{+\infty} \mtr\left(\frac{2L}{L^2+k^2}\right)
     \,h(k)\ \ud k + \frac{\cL}{\pi}\int_{-\infty}^{+\infty}h(k) \ \ud k \\
  &= 2\cL \, \hat h (0) - \frac{1}{2\pi}\int_{-\infty}^{+\infty}
     h(k) \, \frac{\im\mtr S(k)}{k} \ \ud k \ .  
\end{split}
\end{equation*}
Adding the contribution of the Laplace-eigenvalue zero to \eqref{pretrace},
after a rearrangement of the terms the result \eqref{TF1} follows.
\end{proof}

As mentioned in Theorem~\ref{74}, the double sum over periodic orbits in 
\eqref{TF1} does not converge absolutely when $h\in H_r$ with arbitrarily 
small $r\geq0$. In the following we are going to show that an absolutely 
convergent periodic orbit sum can be achieved under sharpened conditions on 
the test function. In order to formulate these conditions we introduce the 
function
\begin{equation}
\label{ldef}
 l(\kappa) := \frac{1}{\kappa}\log(2E) + \frac{2}{\kappa}\arth\Bigl(
 \frac{\kappa}{\lmp}\Bigr) \ ,\quad 0<\kappa<\lmp \ ,
\end{equation}
which attains its unique minimum at some $\sigma\in(0,\lmp)$. Moreover, the 
minimum can be bounded from below as $l(\sigma)\geq (2+\log(2E))/\lmp$ so
that, in particular, $l(\sigma)>2/\lmp$.
\begin{theorem}
\label{thm2}
Let $\Gamma$ be a compact, metric graph with a self adjoint realisation 
$-\Delta(A,B;\vl)$ of the Laplacian, such that the condition 
$l_{\min}>l(\sigma)$ is fulfilled. Furthermore, let $h\in H_r$ be a test 
function with an arbitrary $r\geq\sigma$. Then the following identity
holds:
\begin{equation}
\begin{split}
\label{TF2}
 \sum_{n=0}^{\infty} g _n \, h(k_n) 
 & = \phantom{+} \cL \, \hat{h}(0) + \bigl(g_0 -\tfrac{1}{2}N \bigr) \, h(0)
    -\frac{1}{4\pi}\int_{-\infty}^{+\infty}h(k)\,\frac{\im\mtr S(k)}{k}
    \ \ud k \\
 &  \quad +\sum_{p\in\cP}\left[\bigl(\hat{h}\ast\hat{A}_p \bigr) (l_p) 
    + \bigl(\hat{h}\ast\hat{\overline{A}}_p\bigr)(l_p) \right]\ .
\end{split}
\end{equation}
The quantities appearing are the same as in Theorem~\ref{74}. Here, however,
the sum over periodic orbits converges absolutely.
\end{theorem}
\begin{proof}
The task is to analyse the convergence of the sum over periodic orbits 
in \eqref{TF2} more closely. In Theorem~\ref{74} we only considered the 
convergence of the sum over $l$. This was based on the estimate 
\eqref{lestimate}, which shall now be refined. To this end we refer to
\eqref{ltrace} and \eqref{ltrace2}, and notice that $\mtr[\Lambda(k)U^l(k)]$ 
is a sum consisting of $(2E)^l$ terms 
$s_{i_1 i_2}(k)\dots s_{i_l i_{l+1}}(k)h(k)$, each of which contains a product of 
$l$ factors that are matrix elements of either $S(k)$, $\ui D(\vl)$, or 
$\ui\tfrac{2L}{L^2+k^2}$. These factors are all holomorphic in a strip 
$0\leq\im k <\lmp$ and, following Lemma~\ref{lem:Sextend}, in any strip 
$0\leq \im k\leq\kappa$ with $\kappa<\lmp$, they are bounded from above in 
absolute value by a constant times $(\lmp+\kappa)/(\lmp-\kappa)$. Thus, in
particular, for all $\kappa<\min\{\lmp,r\}$ any product 
$s_{i_1 i_2}(k)\dots s_{i_l i_{l+1}}(k)\,h(k)$ is holomorphic for 
$0\leq\im k\leq\kappa$. Since here we are evaluating Fourier transforms at 
$l_{i_1}+\cdots+l_{i_l}>0$, we can apply a suitable version of the Paley-Wiener 
theorem that is concerned with Fourier transforms of holomorphic functions in 
vertical strips of the upper half plane. This is an obvious, slight variation
of Theorem IX.14 in \cite{Reed:1975}, and implies that there exists 
$C_\kappa>0$ such that
\begin{equation}
\begin{split}
\label{integest}
 \left| \int_{-\infty}^{+\infty}\mtr\bigl[\Lambda(k)U^l(k)\bigr]\,h(k)\ \ud k 
 \right|
 &\leq \sum_{i_1,\dots,i_l =1}^{2E}  \left| \int_{-\infty}^{+\infty}s_{i_1 i_2}(k)
       \dots s_{i_l i_1}(k)\,h(k)\,\ue^{\ui k(l_{i_1}+\cdots+l_{i_l})}\ \ud k 
       \right| \\
 &\leq C_\kappa \left( 2E\,\frac{\lmp+\kappa}{\lmp-\kappa}\,\ue^{-\kappa l_{\min}}
       \right)^l 
 = C_\kappa \ue^{\kappa l \bigl( l(\kappa)-l_{\min}\bigr)}\ , 
\end{split}
\end{equation}
where we made use of the function defined in \eqref{ldef}. We hence conclude 
that under the conditions stated the sum over periodic orbits converges
absolutely. In particular, $l_{\min}>l(\kappa)$ must be fulfilled. The latter 
condition is optimised for $\kappa=\sigma$, at which the function $l(\kappa)$ 
attains its minimum. For the above to hold, the condition $h\in H_r$ with
$r\geq\sigma$ must be satisfied.
\end{proof}
We remark that whenever an amplitude function $A_p(k)$ associated with a
periodic orbit is independent of $k$, as it is always true for non-Robin
boundary conditions, the convolution $\hat{h}\ast\hat{A}_p (l_p)$ degenerates 
into a product $A_p\,\hat{h}(l_p)$. In any case, the condition $r\geq\sigma$
imposed on the test function $h$ ensures that 
$\hat{h}\ast\hat{A}_p(x)=O(\ue^{-\sigma |x|})$, which enables the absolute 
convergence of the periodic orbit sum.
\subsection{The trace of the heat kernel}
The first trace formula of a quantum graph, due to Roth \cite{Roth:1983},
expresses the trace of the heat kernel for the Laplacian with Kirchhoff
boundary conditions in terms of a sum over periodic orbits. This has
recently been extended to all non-Robin boundary conditions by Kostrykin,
Potthoff and Schrader \cite{Schrader:2007}. An application of Theorem~\ref{thm2}
now allows us to produce an appropriate trace formula for all self
adjoint realisations of the Laplacian. In such a case, however, negative 
Laplace eigenvalues $-\kappa_n^2$, with multiplicities $g_n^-$, may occur so
that the trace of the heat kernel,
\begin{equation*}
 \tr\ue^{\Delta t} := \sum_{-\kappa_n^2< 0}g^-_n \, \ue^{\kappa_n^2 t} + 
 \sum_{k_n^2\geq 0}g_n \, \ue^{-k_n^2 t} \ ,\quad t>0 \ ,
\end{equation*}
does not follow immediately from Theorems~\ref{74} and \ref{thm2}. Instead, 
we consider
\begin{equation*}
 \tr_+ \ue^{\Delta t} := \sum_{k_n^2\geq 0}g_n \, \ue^{-k_n^2 t} \ ,\quad t>0 \ .
\end{equation*}
In the trace formula we hence have to choose the (entire holomorphic) function 
$h(k)=\ue^{-k^2 t}$, $t>0$, which is in $H_r$ for all $r\geq 0$. Its Fourier 
transform is $\hat{h}(x)=\tfrac{1}{\sqrt{4\pi t}}\ue^{-x^2/4t}$. We also 
introduce the functions
\begin{equation*}
 a_p (t) := \frac{1}{\sqrt{4\pi t}} \, \int_{-\infty}^{+\infty}\hat{A}_p(l_p-y)\,
 \ue^{-y^2/4t} \ \ud y
\end{equation*}
associated with the periodic orbits on the graph and obtain the following 
statement.
\begin{theorem}
\label{thm3}
Let $\Gamma$ be a compact, metric graph with a self adjoint realisation 
$-\Delta(A,B;\vl)$ of the Laplacian, such that the condition 
$l_{\min}>l(\sigma)$ is fulfilled. Then the following identity holds:
\begin{equation}
\label{TF3}
 \tr_+ \ue^{\Delta t} = \frac{\cL}{\sqrt{4\pi t}} + 
 \bigl(g_0 -\tfrac{1}{2}N \bigr) -\frac{1}{2}\sum_{\alpha=1}^d 
 \frac{\lambda_\alpha}{|\lambda_\alpha|} \, \ue^{\lambda_\alpha^2 t} \,
 \erfc\bigl( |\lambda_\alpha|\sqrt{t} \bigr)
 + 2\sum_{p\in\cP}\re a_p (t)\ .
\end{equation}
Here $d$ is the number of non vanishing eigenvalues $\lambda_\alpha$ of $L$
(counted with their multiplicities) and 
\begin{equation*}
 \erfc(x) = \frac{2}{\sqrt{\pi}}\int_x^{\infty}\ue^{-y^2} \ \ud y \ ,
 \quad x\geq 0 \ ,
\end{equation*}
is the error function complement.  Moreover, as $t\to 0^+$ the trace
of the heat kernel has a complete asymptotic expansion in powers of 
$\sqrt{t}$, whose leading terms read
\begin{equation}
\label{heatexp}
 \tr\ue^{\Delta t} = \frac{\cL}{\sqrt{4\pi t}} + \gamma +O(\sqrt{t}) \ , 
 \quad t\to 0^+ \ .
\end{equation}
Here
\begin{equation*}
 \gamma = g_0 -\frac{N}{2} +\sum_n g_n^- -\frac{1}{2}\sum_j\gamma_{0,j} +
 \frac{1}{2}\sum_j\gamma_{p,j} - \frac{d_+ -d_-}{2} \ ,
\end{equation*}
where $\gamma_{0,j}$ and $\gamma_{p,j}$ are the orders of the finitely many,
non zero, purely imaginary zeros and poles, respectively, of the determinant 
function $F$ and $d_\pm$ denotes the number of positive/negative eigenvalues 
of $L$.
\end{theorem}
\begin{proof}
We first observe that the expression~\eqref{Sdiff} implies 
\begin{equation*}
 \im\mtr S(k) = \mtr\frac{2kL}{L^2 + k^2} = 2k \sum_{\alpha=1}^d 
 \frac{\lambda_\alpha}{\lambda_\alpha^2 + k^2}\ ;
\end{equation*}
then we employ the representation
\begin{equation*}
 \erfc(|x|) = \frac{2|x|}{\pi}\,\ue^{-x^2}\int_{0}^{\infty}
 \frac{\ue^{-y^2}}{y^2 + x^2}\ \ud y \ ,\quad x\in\rz \ .
\end{equation*}
Hence the relation~\eqref{TF3} is an immediate consequence of \eqref{TF2}, 
when the test function $h$ is chosen as indicated above. 

In order to determine the small-$t$ asymptotics we go back to the relation 
\eqref{finK}, in which we use the test function $h(k)=\ue^{-k^2 t}$, $t>0$.
We then deform the contour into $\partial C_{\beta,K}$, with
$\beta>\max\{\lambda_{\max},s\}$, where $s$ from \eqref{boundcond} is such that 
$-s^2$ yields a lower bound on the Laplace spectrum. Thus the contour 
now encloses all non real zeros and poles of the determinant function $F$ and, 
therefore, in this process we pick up contributions from all poles of $F'/F$ 
on the imaginary axis. 

Having to perform the limit $K\to\infty$ with $\beta$ kept fixed, we need
to estimate the contribution coming from the vertical parts of the contour, 
i.e., for $|\re k|=K$ and $\varepsilon<|\im k|<\beta$. Firstly, $F(k)$ is a
polynomial in the matrix entries of $S(k)$ and $T(k)$. The latter are
$\ue^{\ui kl_e}$, whereas the $k$-dependence of the former is given by
$(\lambda_\alpha -\ui k)/(\lambda_\alpha +\ui k)$, see \eqref{5a}. In the
two strips $\varepsilon<|\im k|<\beta$, with neighbourhoods of the poles at
$\ui\lambda_\alpha$ removed, all matrix entries are bounded, and hence $F'(k)$ 
is of polynomial growth in $k$. Secondly, for sufficiently large $|k|$ we 
again approximate $F(k)$ by $F_\infty(k)$, see \eqref{Finftydef}, and write
$F_\infty (k)=\prod_{j=1}^{2E}(1-u_j(k))$. Here $u_1(k),\dots,u_{2E}(k)$ are
the eigenvalues of $S_\infty T(k)$, which satisfy
\begin{equation*}
 |u_j(k)| \leq \| S_\infty \| \, \| T(k) \| \leq \ue^{-\varepsilon l_{\min}}
 < 1 \ ,
\end{equation*}
for all $k$ in the strips $\varepsilon<|\im k|<\beta$. Thus, in these strips,
\begin{equation*}
 |F_\infty (k)| > \bigl( 1- \ue^{-\varepsilon l_{\min}} \bigr)^{2E} >0 \ .
\end{equation*}
Thirdly, when $|K|>\beta$ the factor $\ue^{-t k^2}$ is of the order 
$\ue^{-t K^2}$ in the strips. Hence, the integrand of \eqref{finK} on the 
vertical parts of the contour with $|\im k| >\varepsilon$ is bounded by a 
polynomial times $\ue^{-t K^2}$. As $K\to\infty$, these parts of the contour 
therefore do not contribute to the integral. Hence, from \eqref{finK} we
obtain
\begin{equation*}
\begin{split}
 N + 2\sum_{n=1}^\infty g_n \, \ue^{-k_n^2 t} 
 &= \sum_j\gamma_{p,j}\, \ue^{\kappa_{p,j}^2 t} - \sum_j\gamma_{0,j}\, 
    \ue^{\kappa_{0,j}^2 t} \\
 &\quad +\frac{1}{2\pi\ui}\int_{-\infty}^{+\infty}\left[ \frac{F'}{F}
    (k-\ui\beta)\,\ue^{-(k-\ui\beta)^2 t} -\frac{F'}{F}(k+\ui\beta)\,
    \ue^{-(k+\ui\beta)^2 t} \right]\,\ud k \ . 
\end{split}
\end{equation*}
Here the non real zeros of $F$ are denoted as $\ui\kappa_{0,j}$ and its 
poles as $\ui\kappa_{p,j}$; the respective orders are $\gamma_{0,j}$ and
$\gamma_{p,j}$.

To proceed further we follow the argument leading from \eqref{Unorm} to 
\eqref{logFder1} and \eqref{logFder2}, as well as the subsequent discussion 
of interchanging integration and summation. As we are dealing with large 
$\beta$ instead of small $\varepsilon$, due to \eqref{Sbound3} the estimate 
\eqref{Unorm} is replaced by
\begin{equation*}
\label{Unorm1}
 \| U(k+\ui\beta) \| \leq 
 \max\left\{ 1,\frac{\beta +\lambda_{\max}}{\beta -\lambda_{\max}} \right\} \, 
 \, \ue^{-\varepsilon l_{\min}} \ ,
\end{equation*}
Hence, for $k\in\rz$ and $\beta$ sufficiently large analogous relations to 
\eqref{logFder1} and \eqref{logFder2} are obtained, eventually leading to
\begin{equation*}
\begin{split}
 N + 2\sum_{n=1}^\infty g_n \, \ue^{-k_n^2 t} 
 &= \sum_j\gamma_{p,j}\, \ue^{\kappa_{p,j}^2 t} - \sum_j\gamma_{0,j}\, 
    \ue^{\kappa_{0,j}^2 t} \\
 &\quad + \sum_{l=1}^\infty \frac{1}{2\pi\ui}\int_{-\infty}^{+\infty}\mtr \bigl[ 
    \Lambda(k+\ui\beta)U^l (k+\ui\beta) \bigr] \, \ue^{-(k+\ui\beta)^2 t} 
    \ \ud k \\
 &\quad + \sum_{l=0}^\infty \frac{1}{2\pi\ui}\int_{-\infty}^{+\infty}\mtr \bigl[ 
    \Lambda(k-\ui\beta)U^{-l}(k-\ui\beta) \bigr] \, \ue^{-(k-\ui\beta)^2 t} \ 
    \ud k \ .
\end{split}
\end{equation*}
In the integrals with $l\neq 0$ we now replace $\beta$ by $\beta/\sqrt{t}$, 
$0<t\leq 1$, and change variables from $k$ to $q=k\sqrt{t}$, yielding
\begin{equation*}
 I^\pm_l(t,\beta) := \frac{1}{2\pi\ui\sqrt{t}}\int_{-\infty}^{+\infty}\mtr 
 \bigl[ \Lambda\bigl(\tfrac{1}{\sqrt{t}}(q\pm\ui\beta)\bigr)U^{\pm l}\bigl(
 \tfrac{1}{\sqrt{t}}(q\pm\ui\beta)\bigr) \bigr] \, \ue^{-(q\pm \ui\beta)^2} 
 \ \ud q \ .
\end{equation*}
These integrals can be bounded in analogy to \eqref{integest},
\begin{equation*}
 |I^\pm_l(t,\beta)| \leq \frac{C_\beta}{\sqrt{t}} \left( 2E\,
 \frac{\beta +\sqrt{t}\lambda_{\max}}{\beta -\sqrt{t}\lambda_{\max}}\,
 \ue^{-\beta l_{\min}/\sqrt{t}}\right)^l  \ .
\end{equation*}
Summing over $l\neq 0$ then finally shows that these contributions can be 
estimated as being 
$O\bigl(\tfrac{1}{\sqrt{t}}\ue^{-\beta l_{\min}/\sqrt{t}}\bigr)$.

The term with $l=0$ can be calculated explicitly and yields the same 
contribution to the heat trace as the first term and the sum over 
$\sigma(L)\setminus\{0\}$ on the right-hand side of \eqref{TF3}. 

We add the contributions of negative Laplace eigenvalues and use that
$\erfc(x)$ has a complete asymptotic expansion in $x$, with $\erfc(x)=1+O(x)$,
as $x\to 0$. The expansion \eqref{heatexp} then follows immediately.
\end{proof}
At this point we recall that in the case of non-Robin boundary conditions
$\gamma = g_0 -\tfrac{1}{2}N=\tfrac{1}{4}\mtr S$, which has also been given 
an interpretation as (one half of) a suitable Fredholm index. In this case, 
therefore, the constant term in the small-$t$ asymptotics of the heat kernel 
has a topological meaning. 

Finally, we should like to mention that a suitable Tauberian theorem 
(see, e.g., \cite{Karamata:1931}) allows us to recover Weyl's law
\begin{equation*}
 N(K) \sim \frac{\cL}{\pi}\,K \ , \quad K\to\infty \ ,
\end{equation*}
see also Proposition~\ref{Weylnew}, from the leading term in the 
expansion~\eqref{heatexp}.
\section{Conclusions}
\label{concl}
Our principal goal was to investigate spectra of general self adjoint
realisations of Laplace operators on compact metric graphs, culminating
in proofs of some trace formulae. In this context we achieved to allow for 
a large class of test functions, leading either to absolutely or conditionally
convergent sums over periodic orbits, respectively, representing appropriate 
spectral functions of the Laplacian. 

Previous work on quantum graph trace formulae 
\cite{Roth:1983,KottosSmilansky:1998,Kurasov:2005,Schrader:2007} was 
restricted to Laplacians with non-Robin boundary conditions. As compared to 
these cases there are some modifications we had to take care of. Firstly, 
non-Robin boundary conditions correspond to $k$-independent S-matrices and 
hence do not involve any derivatives of $S(k)$. This is in line with the fact 
that $L=0$, so that in the trace formula the contribution \eqref{Lcontrib} to 
the periodic orbit sum is absent. Moreover, since therefore $\lmp=\infty$, the 
restriction imposed on $l_{\min}$ in the general case is void so that any set
of lengths can be attributed to the edges. What still remains to ensure an 
absolutely convergent periodic orbit sum in the case of non-Robin boundary 
conditions is the single requirement $h\in H_r$ with $r\geq (\log 2E)/l_{\min}$
on the test functions. This implies $\hat h(x)=O(\ue^{-r|x|})$, which in turn 
compensates for the growth in the number of periodic orbits entering the sum
\begin{equation*}
 \sum_{p\in\cP,l_p\leq\ell}A_p \, \hat{h}(l_p) \ ,
\end{equation*}
when $\ell\to\infty$. 

Another property of Laplacians with non-Robin boundary conditions is that 
they are non-negative. This fact is linked to the non-positivity of $L$ which, 
as $L=0$, is trivial. The determinant function~\eqref{23} hence is entire 
holomorphic with only real zeros in the complex non-negative half plane, see 
Proposition~\ref{thm:seceq}.   

As we have shown in Theorem~\ref{74} the condition on the test functions can 
be relaxed to $h\in H_r$ with any $r\geq 0$, so that $h(x)=O(\ue^{-\delta |x|})$
with some (arbitrarily small) $\delta>0$, when one is willing to accept a 
conditionally convergent sum. This has to be understood in the sense given in 
\eqref{TF1}, i.e., where the terms are arranged as a double sum over the 
topological lengths of the orbits and over the periodic orbits of fixed 
topological length. For non-Robin boundary conditions a refined analysis
of convergence had produced even more relaxed conditions to be demanded from
the test functions, see \cite{Winn:2007}.
 
An important application of the trace formula for quantum graphs with non-Robin
boundary conditions was to prove an inverse theorem, very much in the sense of
Kac's famous question `Can one hear the shape of a drum?' \cite{Kac:1966}.
Gutkin and Smilansky \cite{Smilansky:2001} showed that under certain 
conditions, which include the requirement that the edge lengths be rationally 
independent, the Laplace spectrum determines the connectedness and the metric 
structure of a compact metric graph uniquely. In this sense isospectral quantum
graphs are isomorphic. Gutkin and Smilansky made essential use of the trace of 
the wave group,
\begin{equation*}
 \tr\ue^{-\ui t\sqrt{-\Delta}} + c.c. = 2\sum_{n=0}^\infty g_n \, \cos(k_n t) \ ,
 \quad t\neq 0 \ , 
\end{equation*}
which can, in the case of non-Robin boundary conditions, be expressed as a 
sum of $\delta$-singularities at the lengths $l_p$ of periodic orbits.
Searching for these singularities then allows one to first identify the 
geometric length spectrum. In a second step a certain algorithm can be used 
to determine the connectedness and all individual edge lengths. After a slight 
modification this proof can now be taken over to the case of Robin boundary 
conditions almost verbatim. For this purpose one reads off from the trace
formula \eqref{TF2} the distributional identity
\begin{equation*}
\begin{split}
 \sum_{n=0}^\infty g_n \, \cos(k_n t) 
 &= \cL \,\delta(t) + g_0 -\tfrac{1}{2}N - \frac{1}{2}
    \sum_{\alpha=1}^d \frac{\lambda_\alpha}{|\lambda_\alpha|} \, 
    \ue^{-|t\lambda_\alpha|} \\
 &\quad + \sum_{p\in\cP}\re \bigl[ \hat{A}_p (l_p-t)+\hat{A}_p(l_p+t)\bigr]\ .
\end{split}
\end{equation*}
Since, in general, the periodic orbit amplitudes are functions of $k$, there 
are no longer pure $\delta$-singularities present at the lengths of periodic 
orbits. For large $k$ the amplitudes, however, possess the expansions 
\eqref{amplex} with leading terms $a_p^{(0)}$. These do not vanish since they 
stem from the corresponding leading term $S_\infty =\eins - 2P$ of the 
S-matrix. Hence, the Fourier transforms $\hat{A}_p$ of the amplitudes have 
leading singularities of $\delta$-type at the lengths $l_p$ of periodic 
orbits. Therefore, applying the algorithm of Gutkin and Smilanksy to this 
wave trace enables one to identify the graph connectivity and the edge
lengths in the same way as previously.  
 
In summary, almost all spectral properties established so far for Laplacians
on compact metric graphs with non-Robin boundary conditions carry over to
arbitrary self adjoint realisations of the Laplacian. Therefore, there are
many more quantum graphs model available that are suitable for further 
investigations in, e.g., the field of quantum chaos.

\vspace*{0.5cm}
\subsection*{Acknowledgements}
J B would like to thank Stephen Fulling for very helpful discussions.
We are also grateful to an anonymous referee for useful hints.
\appendix

\newpage

{\small
\bibliographystyle{amsalpha}

\providecommand{\bysame}{\leavevmode\hbox to3em{\hrulefill}\thinspace}
\providecommand{\MR}{\relax\ifhmode\unskip\space\fi MR }
\providecommand{\MRhref}[2]{%
  \href{http://www.ams.org/mathscinet-getitem?mr=#1}{#2}
}
\providecommand{\href}[2]{#2}

\end{document}